\documentclass[a4paper, 11pt]{article}
\pdfoutput=1

\usepackage[dvipsnames]{xcolor}
\usepackage{comment}
\usepackage{jcappub}
\usepackage{graphicx}
\usepackage{booktabs}
\usepackage{siunitx}
\usepackage{soul}
\usepackage{lineno}

 \newcommand{\rhodm}{\rho_\mathrm{DM}}

\title{Ultralight dark matter searches at the sub-Hz frontier with atom multigradiometry}

\author[a]{Leonardo Badurina,}
\author[b]{Valerie Gibson,}
\author[a]{Christopher McCabe}
\author[b]{and Jeremiah Mitchell}

\affiliation[a]{Theoretical Particle Physics and Cosmology Group, Department of Physics, King's College London, Strand, London, WC2R 2LS, United Kingdom}
\affiliation[b]{Cavendish Laboratory, Department of Physics, University of Cambridge, Cambridge, CB3 0HE, United Kingdom}

\emailAdd{leonardo.badurina@kcl.ac.uk}

\abstract{
Single-photon atom gradiometry is a powerful experimental technique that can be employed to search for the oscillation of atomic transition energies induced by ultralight scalar dark matter (ULDM). 
In the sub-Hz regime, the background is expected to be dominated by gravity gradient noise (GGN), which arises as a result of mass fluctuations around the experiment. 
In this work, we model the GGN as surface Rayleigh waves, and we construct a likelihood-based analysis that consistently folds GGN into the sensitivity estimates of vertical atom gradiometers in the frequency window between $10^{-3}$\,Hz and 1\,Hz.
We show that in certain geological settings GGN can be significantly mitigated when operating a multigradiometer configuration, which consists of  three or more atom interferometers in the same baseline.
Multigradiometer experiments, such as future versions of AION and MAGIS-100, have the potential to probe regions of scalar ULDM parameter space in the sub-Hz regime that have not been excluded by existing experiments.
}

 \keywords{KCL-PH-TH/2022-29, AION-REPORT/2022-3}

\begin{document}
\maketitle

\flushbottom

\section{Introduction}

Atom interferometry is a rapidly-developing and powerful experimental technique that can be employed in different configurations for a wide variety of precision measurements (see e.g.,~\cite{Peters2001, Canuel2006,Rosi2014,Parker2018,Asenbaum2020,Stray2022}).
For example, two spatially-separated atom interferometers (AIs) that are referenced by the same laser sources can be operated as a gradiometer to shed light on different topics in fundamental physics.
Specifically, owing to their exquisite sensitivity to timings, accelerations and changes in atomic structures, atom gradiometers could be used for: the detection of gravitational waves (GW) in the unexplored `mid-frequency band'~\cite{Dimopoulos:2007cj, Graham:2012sy,Ellis:2020lxl,Badurina:2021rgt}; the search of violations of the universality of free-fall arising from static and time-dependent fifth forces acting on matter~\cite{Graham:2015ifn}; the detection of sub-GeV dark matter through quantum decoherence effects~\cite{Riedel:2016acj, Du:2022ceh}; and the search for oscillating atomic transition energies induced by scalar ultralight dark matter (ULDM)~\cite{Arvanitaki:2016fyj, Badurina:2021lwr}, which will be the focus of this paper. 

It has been shown that several terrestrial long-baseline atom gradiometer and atom interferometer network projects, including AION~\cite{Badurina:2019hst}, MAGIS-100~\cite{MAGIS-100:2021etm}, MIGA~\cite{Canuel:2017rrp}, ELGAR~\cite{Canuel:2019abg}, ZAIGA~\cite{Zhan:2019quq}, and space-borne proposals, such as STE-QUEST~\cite{Aguilera:2013uua} and AEDGE~\cite{Bertoldi:2019tck}, could probe regions of scalar ULDM parameter space that have not been excluded by other experiments, such as atomic clocks~\cite{Kennedy:2020bac}, torsion balances~\cite{Wagner:2012ui} and the MICROSCOPE experiment~\cite{PhysRevLett.129.121102}. For example, compact atom gradiometers such as AION-10, where the baseline is $\sim\SI{10}{m}$, offer the possibility of detecting scalar ULDM with a mass of approximately $\SI{e-15}{eV}$~\cite{Badurina:2021lwr}, which corresponds to a signal oscillating at a frequency of approximately $\SI{1}{Hz}$. Longer baseline gradiometers like AION-km~\cite{Badurina:2019hst} and MAGIS-km~\cite{MAGIS-100:2021etm} could potentially detect ULDM across an even larger frequency window: from $\sim \SI{e-3}{Hz}$ to $\sim \SI{e3}{Hz}$, corresponding to ULDM masses between $\sim \SI{e-17}{eV}$ and $\sim \SI{e-11}{eV}$~\cite{Badurina:2021rgt}.

The projected reach of these experiments is ultimately limited by fundamental noise sources.
Vertical gradiometers such as AION and MAGIS-100 are designed to reach the atom shot-noise limit above $\sim\SI{1}{Hz}$~\cite{MAGIS-100:2021etm}, but would suffer from gravity gradient noise (GGN) at lower frequencies~\cite{Arvanitaki:2016fyj}. This type of phase noise arises as a result of mass density fluctuations of the ground and atmosphere~\cite{Harms:2015zma}, which perturb the local gravitational potential around the atom clouds. In the absence of a well-modelled GGN signal for vertical atom interferometers, previous projections were either abruptly interrupted at $\sim \SI{0.3}{Hz}$~\cite{Arvanitaki:2016fyj,Badurina:2019hst, Badurina:2021rgt}, or arbitrarily extended to lower frequencies by assuming atom shot noise only~\cite{Badurina:2021lwr}.

On the basis of the seminal GGN studies for optical gravitational wave detectors~\cite{hughes_seismic_1998, Beccaria:1998ap}, Refs.~\cite{Baker:2012ck,Vetrano:2013qqa,Harms:2013raa} provided the first characterisation of the impact of GGN on a vertical km-long gradiometer. Within the context of gravitational wave searches, Refs.~\cite{Baker:2012ck,Vetrano:2013qqa,Harms:2013raa} concluded that a pair of coupled atom interferometers provides a sizable, though not dramatic, reduction of the GGN between 0.1~Hz and 1~Hz compared to large scale optical interferometers. 
Subsequently, Ref.~\cite{MIGAconsortium:2019efk} showed how GGN from seismic activity and pressure variations dominates the sub-Hz background in the MIGA network of atom interferometers.
In light of the fact that GGN is characterised by an exponential profile that decays with radial distance from the Earth's surface, while a gravitational wave signal scales linearly with the length of the baseline,
Ref.~\cite{Mitchell_2022} showed that GGN could in principle be distinguished from a gravitational wave signal by placing several interferometers along the baseline.  Furthermore, Ref.~\cite{Mitchell_2022} demonstrated that increasing the number of interferometers along the baseline provides better suppression of the GGN signal, in agreement with Ref.~\cite{MIGAconsortium:2019efk}, which found a similar suppression for a horizontal configuration.  
These previous studies~\cite{Baker:2012ck,Vetrano:2013qqa,Harms:2013raa, MIGAconsortium:2019efk, Mitchell_2022} have focused on GGN effects in the context of gravitational wave searches but have not considered the impact of GGN on generic physics searches, such as ULDM-dedicated runs, using robust statistical techniques. 

In this paper we develop the formalism to consider the impact of GGN on ULDM searches with a long-baseline vertical gradiometer. Such a formalism will be of paramount importance to characterise the sensitivity of gradiometers, such as long-baseline versions of AION and MAGIS-100, to scalar ULDM across the frequency range from $\SI{e-3}{Hz}$ to~$\SI{1}{Hz}$. In contrast to previous studies, we utilise a likelihood-based analysis, which is more commonly used in the particle-physics community, so that the impact of GGN on ULDM searches can be placed on a more robust statistical footing. In our analysis, we simultaneously account for an ULDM signal and a background dominated by atom shot noise and underground GGN. We also extend previous studies by robustly quantifying the extent to which GGN underground can be mitigated in a multigradiometer experiment that consists of three or more coupled atom interferometers along the same baseline.

This paper is structured as follows. In section~\ref{sec:AtomGrad} we briefly review how atom gradiometers can be used for fundamental physics searches, and we define the key features of an atom multigradiometer. In section~\ref{sec:StatsFormalism} we present the statistical properties of the ULDM signal and backgrounds in an atom multigradiometer to then construct in section~\ref{sec:AnalysisMultiGrad} the appropriate likelihood analysis for setting projected upper limits. In section~\ref{sec:ULDMGGNBackground} we discuss the impact of GGN on ULDM searches, firstly in the context of a single gradiometer, before proceeding to show how a multigradiometer configuration can suppress GGN in different configurations. Finally, we provide a discussion of these results in section~\ref{sec:Discussion}. Appendices~\ref{app:ULDMphase}-\ref{app:AsimovTestStatistics} provide further details and derivations of calculations to support the results in the main body of the paper.

\section{Atom gradiometry for precision measurements}\label{sec:AtomGrad}

An atom interferometer is an experiment that compares the  
phase between coherent, spatially delocalised quantum superpositions of atom clouds~\cite{Abend:2020djo}. Experimentally, this can be achieved by constructing a two-level system, composed of an excited state $| e \rangle$ and a ground state $| g \rangle$ with energy separation $ \omega_A$.\footnote{Unless stated otherwise, we use natural units and set $\hbar=c=1$.} The electronic transitions $| e \rangle \leftrightarrow| g \rangle$, as well as the degree of quantum superposition and their physical separation, can be controlled by using a system of lasers. In this paper, we consider single-photon atom interferometers, which operate with atoms whose two-level system is controlled by single-photon transitions, as is the case for the $^{87}\mathrm{Sr}$ isotope: the ground state corresponds to $5\mathrm{s}^2\, ^{1}\mathrm{S}_0$, the excited state to $5\mathrm{s}5\mathrm{p}\, ^{3}\mathrm{P}_0$, and $\omega_A\approx2.697\times10^{15}~\mathrm{rad}/\mathrm{s}$ (corresponding to a wavelength $\lambda_A\approx \SI{698}{nm}$). The superposition of ground and excited states is controlled via $\pi/2$-pulses (i.e.\ beam-splitter pulses), after which the excited state experiences a momentum kick in the vertical direction that allows for the two atomic states to spatially separate during propagation. The atom's internal state and linear momentum can be reversed using $\pi$-pulses (i.e.\ mirror pulses).\footnote{A $\pi/2$-pulse is defined as a pulse of resonant radiation that interacts with the atoms for a time $\pi/(2 \Omega)$, where $\Omega$ is the Rabi frequency. A $\pi$-pulse interacts with the atoms for a time $\pi/\Omega$.}

\begin{figure*}[t!]
    \centering
    \includegraphics[width=1\textwidth]{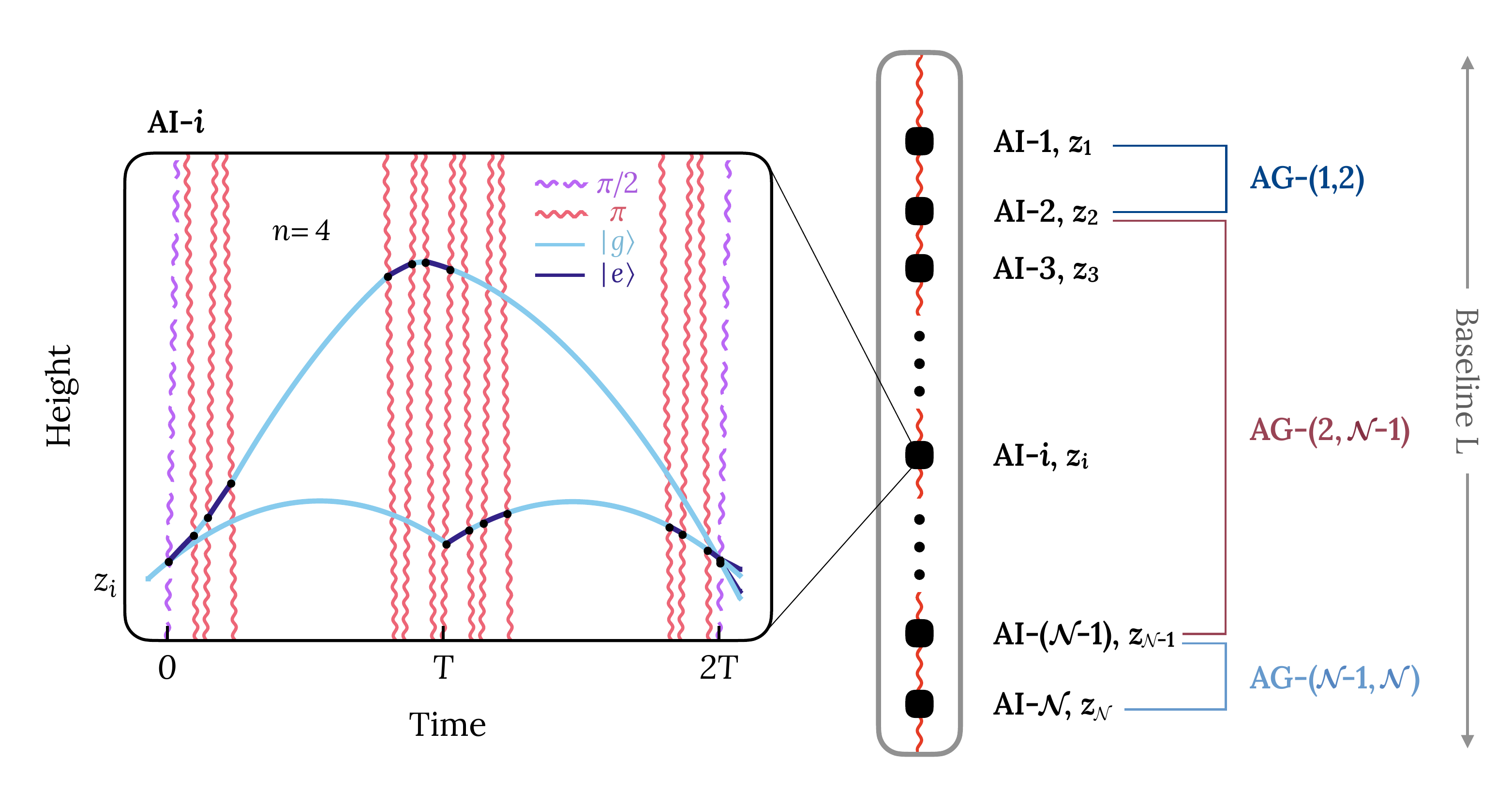}
    \caption{Schematic representation of an atom multigradiometer. The left panel shows the not-to-scale spacetime diagram of the $i^\mathrm{th}$ atom interferometer's sequence with $n=4$ LMT kicks, where the atom starts in the ground state at position~$z_i$. The atom's excited ($| e \rangle$) and ground ($| g \rangle$) states are shown in blue and cyan, respectively. $\pi/2$- and $\pi$-pulses are displayed as wavy lines in fuchsia (dashed) and red (solid), respectively. Atom-light interactions are indicated with black dots. The left panel is to be interpreted as a blow-up of each of the right panel's black squares, which represent the atom interferometers along the vertical baseline of length $L$ and are labeled by $\text{AI-}i$, where $i = \{1,2,...,\mathcal{N}\}$. The vertical position of the $i^\mathrm{th}$ interferometer is labeled by $z_i$. The origin of our coordinate system is at the Earth's surface at the centre of the interferometer shaft, so the shorted vertical distance from the surface in this underground detector is $z_1$. Atom gradiometers are represented by braces connecting spatially separated pairs of atom interferometers. The atom gradiometer composed of the $i^\mathrm{th}$ and $j^\mathrm{th}$ atom interferometer is labeled as~$\text{AG-}(i,j)$.
    }
    \label{fig:AI_layout}
\end{figure*}

To increase the separation between atom clouds, large momentum transfer (LMT) mirror pulses have been proposed~\cite{Hogan2009}. With our convention, which follows Ref.~\cite{Badurina:2021lwr}, a sequence characterised by $n$ LMT `kicks' features $4n-1$ pulses in total, of which two are $\pi/2$-pulses and the remaining $4n-3$ are $\pi$-pulses. These are distributed such that at the beginning and the end of the sequence, there are ($n-1$) $\pi$-pulses (the `beamsplitter sequence'), while around time $T$, known as the interrogation time, there are ($2n-1$) $\pi$-pulses (the ‘mirror sequence’).
With our labelling convention, the $\pi$/2-pulses are emitted at times 0 and $2T$. An example of a sequence with $n=4$ LMT kicks is shown in the left panel of Fig.~\ref{fig:AI_layout}. For clarity, the time axis in Fig.~\ref{fig:AI_layout} is not shown to scale: the typical interrogation time $T$ is $\mathcal{O}(\mathrm{ s})$ while the time between individual pulses is typically $\mathcal{O}(100\,\mathrm{ n s})$~\cite{Rudolph:2019vcv,wilkason_atom_2022}.

In the final stage of the interferometric sequence, each atom is in a superposition of two velocity states, which thus spatially separate. After an appropriate drift time, the two paths spatially separate allowing for the determination of the atom populations via atom fluorescence and point source atom interferometry~\cite{rocco_fluorescence_2014,dickerson_multiaxis_2013}. The probability that an atom will be found in a particular output port depends on the relative phase $\Phi$ acquired along the two paths of the atom interferometer and can be measured from the number of atoms in the excited or ground states, $N_e$ and~$N_g$ respectively, as 
\begin{equation}
\frac{N_e-N_g}{N_g+N_e} =  C \cos{\Phi}  \, , 
\label{eq:PhaseDefinition}
\end{equation}   
where $C \leq 1$, which is known in the literature as \textit{constrast}, characterises the amplitude of these oscillations~\cite{Le_Gou_t_2008, PhysRevA.47.3554}. The contrast is maximised when the relative displacement between the wave packets vanishes at the end of the interferometric sequence. Hence, knowing the value of $C$, the total phase difference $\Phi$ can be determined by using the data collected at the interferometer's output ports and Eq.~\eqref{eq:PhaseDefinition}.

The phase difference $\Phi$ results from both the free-fall evolution of the quantum state along each path, and the local phase of the laser which is imprinted onto the atom during each atom-light interaction. Since the interferometer sequence compares the motion of the atom to the reference frame defined by the laser pulses, $\Phi$ is highly sensitive to inertial forces acting on the atom during the interferometer sequence. In this sense, atom interferometers can be used as accelerometers. Additionally, since the interferometer sequence is composed of paths during which the atomic state changes from the ground to the excited state, $\Phi$ is also highly sensitive to the constants that govern atomic transitions. In this sense, atom interferometers can be understood as atomic clocks.

Two spatially separated atom interferometers that are referenced by the same laser pulse via single-photon atomic transitions can be operated as a gradiometer~\cite{Graham:2012sy}. The distinctive signal of a coherently oscillating ULDM field~\cite{Arvanitaki:2016fyj} and GW passing through the experiment~\cite{Graham:2015ifn} can be tested with a gradiometer by finding the difference between the phases measured by two atom interferometers.  Vertical gradiometers like MAGIS-100 and AION, as opposed to horizontal configurations envisaged for MIGA, ELGAR and ZAIGA, consist of interferometers that are positioned along the direction of the Earth's gravitational acceleration. A key advantage of the vertical gradiometer experimental set-up over a lone single-photon atom interferometer relies on the possibility of effectively attenuating laser noise. By operating the same laser pulse between both interferometers, the laser noise affecting the experiment's sensitivity cancels in a differential measurement~\cite{Graham:2012sy}.

\subsection{Atom multigradiometry}\label{sec:AtomMultiGrad}

We define an experiment consisting of $\mathcal{N} \geq 2$ atom interferometers that are referenced by the same set of lasers and located within the same baseline of length~$L$ as an \textit{atom multigradiometer}. This is to be contrasted with two atom gradiometers operating in different baselines, irrespective of their synchronisation. A schematic representation of this setup is shown in the right panel of Fig.~\ref{fig:AI_layout},
where the black squares represent the~$\mathcal{N}$ AIs along the baseline.
We will assume that the AIs can be situated at any point along the baseline and, in particular, that they do not have to be equally spaced.

An atom gradiometer (AG) measurement consists of taking the difference between the phases collected by two AIs. Figure~\ref{fig:AI_layout} shows an example of three AG measurements: using the first and second AIs; the $(\mathcal{N}-1)^{\mathrm{th}}$ and $\mathcal{N}^{\mathrm{\,th}}$ AIs; and the second and $(\mathcal{N}-1)^{\mathrm{th}}$ AIs. In total there are~$\mathcal{N}(\mathcal{N}-1)/2$ unique gradiometer measurements that can be performed. It then follows that a multigradiometer experiment employing two atom interferometers corresponds to a single atom gradiometer, as considered in previous works (see e.g.,~\cite{Graham:2016plp, Arvanitaki:2016fyj,Badurina:2019hst,Badurina:2021lwr}).\footnote{The set-up envisaged here differs from the interferometer array planned for the MIGA and ELGAR experiments, where $\mathcal{N}$ independent gradiometer measurements are performed on two {\it horizontal} baselines, which feature $2\mathcal{N}$ interferometers each~\cite{Chaibi:2016dze, MIGAconsortium:2019efk}.
}
Since the interferometers are located within the same baseline and are referenced by the same laser, the AIs will be characterised by the same interrogation time and number of LMT kicks. To simplify the subsequent analysis, we will assume that the launch velocity and atom number are identical in all of the AIs. 

\section{Characterising signal and noise in an atom gradiometer} \label{sec:StatsFormalism}

Interactions between {\it scalar} ULDM and electrons or photons can lead to oscillations in fundamental constants (see e.g.~\cite{Arvanitaki:2014faa, Stadnik:2014tta,Stadnik:2015kia}). These ULDM-induced oscillations in turn alter atomic transition energies, and can be searched for with atom interferometer experiments~\cite{Arvanitaki:2016fyj, Badurina:2021lwr}.
In this section, we characterise the
properties of the ULDM signal and the backgrounds considered in this work, namely GGN from seismic effects and atom shot noise, as recorded by an atom multigradiometer operating in a vertical configuration. While an interferometer's atom shot noise is an irreducible noise source and is uncorrelated between detectors, the ULDM signal and the GGN background will induce non-trivial cross-correlations between pairs of interferometers that can be exploited to improve the sensitivity to low frequency ULDM signals.

An atom interferometer measures a time-dependent signal by collecting phase differences over a campaign of duration~$T_\mathrm{int}$: we refer to this as the integration time and it is measured with respect to the first phase measurement. Measurements are made at a sampling rate~$1/\Delta t$, where~$\Delta t$ is the temporal separation between successive measurements. We assume that~$\Delta t$ takes the same value over the measurement campaign. The integration time satisfies the relation $T_\mathrm{int} = (N-1)\Delta t$, where $N$ is the number of measurements. Since $T_\mathrm{int}$ is assumed to be $\mathcal{O}(\mathrm{years})$, and $\Delta t$ is typically $\mathcal{O}(1~\mathrm{s})$, we will use the approximation $N \approx T_\mathrm{int}/\Delta t$.
 
We define the phase measured by the~$i$\textsuperscript{th} interferometer ($\text{AI-}i$) at the end of a sequence that starts at time $m\Delta t$~as
\begin{equation}
\Phi^{(i)}_m = \Phi^{(i)}_{\mathrm{Signal},m} + \Phi^{(i)}_{\mathrm{Noise},m} \, ,
\end{equation}
where $i \in \{1,2,...,\mathcal{N}\}$, $m \in \{0,1,...,N-1\}$, $\Phi^{(i)}_{\mathrm{Signal},m}$ is the signal phase and~$\Phi^{(i)}_{\mathrm{Noise},m}$ is the noise or background contribution. The gradiometer phase between the $i$\textsuperscript{th} and $j$\textsuperscript{th} AI, which will be referred to as $\text{AG-}(i,j)$, is defined~as
\begin{equation}
\Phi^{(i,j)}_m = \Phi^{(i)}_m-\Phi^{(j)}_m = \Phi^{(i,j)}_{\mathrm{Signal},m} + \Phi^{(i,j)}_{\mathrm{Noise},m} \;,
\end{equation}
where $i,j \in \{1,2,...,\mathcal{N}\}$,  $j > i$, and the gradiometer signal and noise contributions are respectively defined~as
\begin{align}
\Phi^{(i,j)}_{\mathrm{Signal},m} &= \Phi^{(i)}_{\mathrm{Signal},m} - \Phi^{(j)}_{\mathrm{Signal},m} \, , \\
\Phi^{(i,j)}_{\mathrm{Noise},m} &= \Phi^{(i)}_{\mathrm{Noise},m} - \Phi^{(j)}_{\mathrm{Noise},m}\;.
\end{align}

For ULDM searches, $\Phi^{(i,j)}_{\mathrm{Signal},m}$ would correspond to the differential ULDM phase recorded by the $i$\textsuperscript{th} and $j$\textsuperscript{th} AI. Instead, for GW searches, the phase would correspond to the GW gradiometer signal. In either case, the phase noise $\Phi^{(i,j)}_{\mathrm{Noise},m}$ captures all background contributions, including static effects, such as the Earth's gravitational field experienced by the atoms in free-fall, and time-dependent effects, such as~GGN and atom shot noise. Because of the nature of the signal of interest, in this paper we shall focus on atom shot noise and GGN, as these are expected to be the dominant noise sources for long-baseline interferometry at high and low frequencies, respectively. 

\subsection{Ultralight dark matter signal}

ULDM can be modeled as a temporally and spatially oscillating non-relativistic classical field. This follows from its high occupation number, small mean velocity and velocity dispersion, which is characteristic of dark matter in the Milky Way~\cite{Hui:2021tkt}. We assume that the ULDM field can be decomposed into Fourier modes with frequencies sampled from the ULDM speed distribution, and random phases sampled from a uniform distribution between 0 and $2\pi$. We assume a Maxwellian distribution of ULDM speeds according to the Standard Halo Model (SHM)~\cite{Lewin:1995rx, Drukier:1986tm} given by
\begin{equation}\label{eg:fSHM}
\begin{split}
    f_{\rm{DM}} (v) &=\frac{v}{ \sqrt{2 \pi} \sigma_v v_{\mathrm{obs}}} e^{-\left(v+v_{\mathrm{obs}}\right)^{2} / (2 \sigma_v^2)} \left(e^{4 v v_{\mathrm{obs}} / (2 \sigma_v^2)}-1\right)\, ,
    \end{split}
\end{equation}
where $\sigma_v$ is the velocity dispersion which is set, at the solar position, by the value of the local standard of rest   $v_0= \sqrt{2 }\sigma_v \approx \SI{238}{km/s}$, and $v_\mathrm{obs} \approx \SI{252}{km/s}$ is the average speed of the Earth relative to the halo rest frame~\cite{Baxter:2021pqo}. Although the DM speed distribution is characterised by a cut-off at the escape velocity, $v_{\rm{esc}} \sim \SI{800}{km/s}$ in the Earth's frame, and may feature additional, anisotropic components (e.g.,~\cite{OHare:2018trr,Evans:2018bqy,Necib:2019zbk,OHare:2019qxc}), the simple form in Eq.~\eqref{eg:fSHM} with no cut-off and no anisotropies is sufficient for our discussion.\footnote{The untruncated SHM sets constraints on the ULDM-SM couplings that are $\lesssim0.1\%$ weaker than those determined using the truncated SHM. 
As shown for axion haloscopes, which also analyse the ULDM signal in the frequency domain, the truncated SHM sets constraints on the ULDM-SM couplings that are $\sim 8\%$ weaker than those determined using the truncated SHM++~\cite{OHare:2018trr}, which considers different speed parameters and a strongly radially anisotropic sub-component due to the Gaia Sausage. These modified speed parameters are largely responsible for the size of this correction.}

The atom interferometer's integration time sets the frequency resolution, $\Delta f= 1/T_\mathrm{int}$, which in turn sets the ULDM speed resolution $\Delta v$, since $\Delta f = m_\phi v \Delta v/2\pi$, where $m_\phi$~is the mass of the scalar field. The former relation follows from the properties of the discrete Fourier transform of the data, while the latter follows from the ULDM's kinetic energy definition.
In an experiment capable of resolving the ULDM speed resolution to an accuracy of~$\Delta v$, the scalar ULDM field can be written as
\begin{equation}
\phi(t) = \frac{\sqrt{\rhodm}}{m_\phi}\sum_{a}{\alpha_a \sqrt{F_\mathrm{DM}(v_a)} \cos{\left [\omega_a t + \theta_a \right]}} \, , \label{eq:full-field}
\end{equation}
where $\rhodm$ is the local dark matter density to which we assign the value $0.3~\mathrm{GeV}/\mathrm{cm}^3$, $\omega_a \simeq m_\phi(1+v_a^2/2)$ is the angular frequency of the ULDM wave for a speed~$v_a$,  
\begin{equation}
F_\mathrm{DM}(v_a) = \int_{v_a - \Delta v/2}^{v_a + \Delta v/2} dv \, f_\mathrm{DM}(v)\;,
\label{eq:F_DM}
\end{equation}
and the sum is over experimentally-resolvable speeds which are indexed by the variable $a$~\cite{Foster:2017hbq}.
The variable~$\theta_a\in [0,2\pi)$ is a random phase, which we assume to be uniformly distributed, while $\alpha_a$ is Rayleigh distributed with $\langle \alpha_a^2 \rangle  = 2$; its probability density function is given by
\begin{equation}
    P(\alpha_a) = \alpha_a \exp \left ( - \frac{\alpha_a^2}{2} \right ) \, .
    \label{eq: Rayleigh distribution}
\end{equation}
Importantly for the subsequent discussion, the random phase~$\theta_a$ and the Rayleigh variable~$\alpha_a$ are independent~\cite{Foster:2017hbq}.

Eq.~\eqref{eq:full-field} correctly captures the field's behaviour in the limit of long and short integration time with respect to the ULDM's coherence time, $\tau_c = 2\pi/(m_\phi v_0^2)$. Since $f_\mathrm{DM}(v)$ is non-vanishing between zero and $v_\mathrm{esc}$, and $v_\mathrm{esc}\gtrsim v_{0}$, Eq.~\eqref{eq:F_DM} is equal to one for $v \sim v_0$ and $\Delta v \gtrsim v_0$ which, from the relation $\Delta v = 2\pi/m_\phi v T_\mathrm{int}$, implies that $T_\mathrm{int} \lesssim \tau_c$. Hence, in this limit, Eq.~\eqref{eq:full-field} is independent of the speed distribution and the DM wave is described in terms of a single Fourier component~\cite{Centers:2019dyn}. If $\Delta v \ll v_0$, then the signal is expressed in terms of a sum of Fourier modes of different frequencies. From the relation $\Delta v = 2\pi/m_\phi v T_\mathrm{int}$, this implies that $T_\mathrm{int} \gg \tau_c$. In this limit, $F_\mathrm{DM}(v_a) \approx f_\mathrm{DM}(v_a) \Delta v$, such that Eq.~\eqref{eq:full-field} is in agreement with Refs.~\cite{Foster:2017hbq, Foster:2020fln}, which consider the limit $T_\mathrm{int} \rightarrow \infty$.

In Eq.~\eqref{eq:full-field} we neglected the spatial variation of the ULDM wave. We justify this by noting that on average the wave vector is suppressed by a factor of $v \sim 10^{-3}$ relative to the angular frequency. Furthermore, the lengths of the baselines that we consider are significantly smaller than the ULDM's de Broglie wavelength, such that any spatially-dependent contribution to the phase of the DM wave is highly subdominant. Indeed, the frequency range of interest corresponds to ULDM masses between \SI{e-17}{eV} and \SI{e-10}{eV}, which would be associated with de Broglie wavelengths of length \SI{e11}{km} and \SI{e4}{km}, respectively; for reference, the experimental baselines considered here are no longer than 1~km.

In this work, we consider a linearly-coupled scalar ULDM field that has interactions governed by the Lagrangian 
\begin{equation}
\mathcal{L}_{\phi} \supset \phi(t) \sqrt{4 \pi G_{N}} \left[\frac{d_{e}}{4 e^{2}} F_{\mu \nu} F^{\mu \nu} -d_{m_{e}} m_{e} \overline{\psi}_e\psi_e \right]  \, .
\end{equation}
Here, $d_e$ and $d_{m_e}$ are dimensionless and parameterise the coupling strength relative to the Planck mass $1/\sqrt{4\pi G_N}$, where  $G_N$ is Newton's gravitational constant.
These interactions will induce oscillations in an atom's transition frequency, which in turn will lead to a phase shift in an atom gradiometer experiment (see e.g.,~\cite{Arvanitaki:2014faa, Stadnik:2014tta, Stadnik:2015kia}). Specifically, the phase shift induced by linearly-coupled scalar ULDM for the gradiometer $\text{AG-}(i,j)$, where the two AIs are separated by a distance $\Delta z^{(i,j)} = z_i-z_j$, can be parametrised~as
\begin{equation}\label{eq:ULDMphase}
   \Phi^{(i,j)}_{\mathrm{DM},m} = \frac{\Delta z^{(i,j)}}{L} \sum_{a} \alpha_a  \sqrt{F_\mathrm{DM}(v_a)} \, A_a \cos \phi_{a,m}\;.
\end{equation}
Here, $\phi_{a,m} \supset \omega_a m\Delta t + \theta_a$ is the phase of the DM wave at the end of the sequence. The amplitude $A_a$ is dependent on the experimental variables $n$, $T$, and $L$ characterising the interferometric sequence, and on phenomenological parameters, the most important of which is the ULDM-Standard Model coupling strength $d_\phi \in \{d_e, d_{m_e}\}$. In the limit $n L \ll T$ and $\omega_a n L \ll 1$, which are always satisfied in planned interferometers~\cite{Badurina:2019hst, MAGIS-100:2021etm}, 
\begin{equation}
    A_a \propto \frac{d_\phi \, n \, L}{\omega_a} \sin^2 \left(\frac{\omega_{a}T}{2} \right)  \, .   
    \label{eq: ULDM amp}    
\end{equation}
The complete expression for $A_a$ is given in Appendix~\ref{app:ULDMphase}. 

Since $\alpha_a$ and $\theta_a$ are independent, over long timescales the signal is characterised by a vanishing expectation value,
\begin{equation}
\begin{aligned}
\Big \langle \Phi^{(i,j)}_{\mathrm{DM}, m} \Big \rangle &= 0 \, ,
\end{aligned}
\label{eq:ULDMexpectation}
\end{equation}
and a covariance given by
\begin{equation}
\begin{split}
\Big \langle \Phi^{(i,j)}_{\mathrm{DM}, m} \Phi^{(i',j')}_{\mathrm{DM}, m'}\Big \rangle 
& =\frac{\Delta z^{(i,j)}}{L} \frac{\Delta z^{(i',j')}}{L} \sum_{a} F_\mathrm{DM}(v_a)   A_a^2 \cos (\omega_a \Delta t(m-m')) \, .
\end{split}
\label{eq:ULDMcovariance}
\end{equation}
Both of these expressions follow from the statistical properties of the amplitude and random phase. A complete derivation of Eqs.~\eqref{eq:ULDMexpectation} and \eqref{eq:ULDMcovariance} is provided in Appendix~\ref{app:ULDMphaseStatisticalProperties}.

\subsection{Gravity gradient noise from seismic effects} \label{sec:GGN}

\begin{figure}[t!]
\centering
\includegraphics[width=0.99\columnwidth]{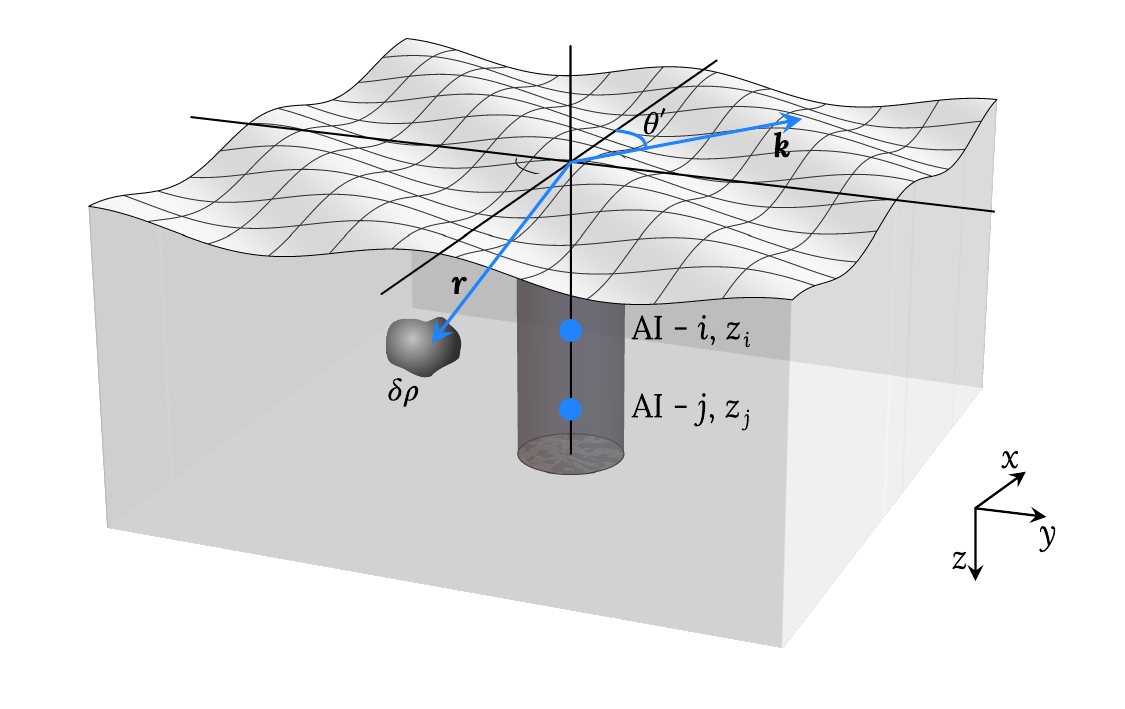}
\caption{Schematic representation of the seismic field for a fundamental Rayleigh mode with angular frequency $\omega_a$ travelling horizontally in the direction $\hat{\mathbf{k}}$. The Rayleigh wave will induce underground density variations $\delta \rho$ at positions labelled by $\mathbf{r}$. The density variations induce perturbations in the gravitational potential, which serve as a noise source for the $i^\mathrm{th}$ interferometer AI-$i$.}
\label{fig: GGN}
\end{figure}

Time-dependent perturbations to the gravitational potential around the free-falling atom clouds cause a phase shift to build up between two paths of an atom interferometer. 
These perturbations in the gravitational potential originate from terrestrial density fluctuations close to each atom interferometer, as illustrated in Fig.~\ref{fig: GGN}, and give rise to a noise source commonly referred to as gravity gradient noise (GGN). 
For the experimental configuration and frequency range considered in this paper, the dominant source of GGN is expected to consist of ground density perturbations induced by ambient horizontally-propagating seismic waves that are confined near the Earth's surface by horizontal geological strata~\cite{hughes_seismic_1998, Harms:2015zma}.
The modes responsible for these density fluctuations are referred to as Rayleigh modes, and they are generated at horizontal discontinuities at strata interfaces, including the Earth's surface.

\begin{figure}[t!]
\centering
\includegraphics[width=0.75\columnwidth]{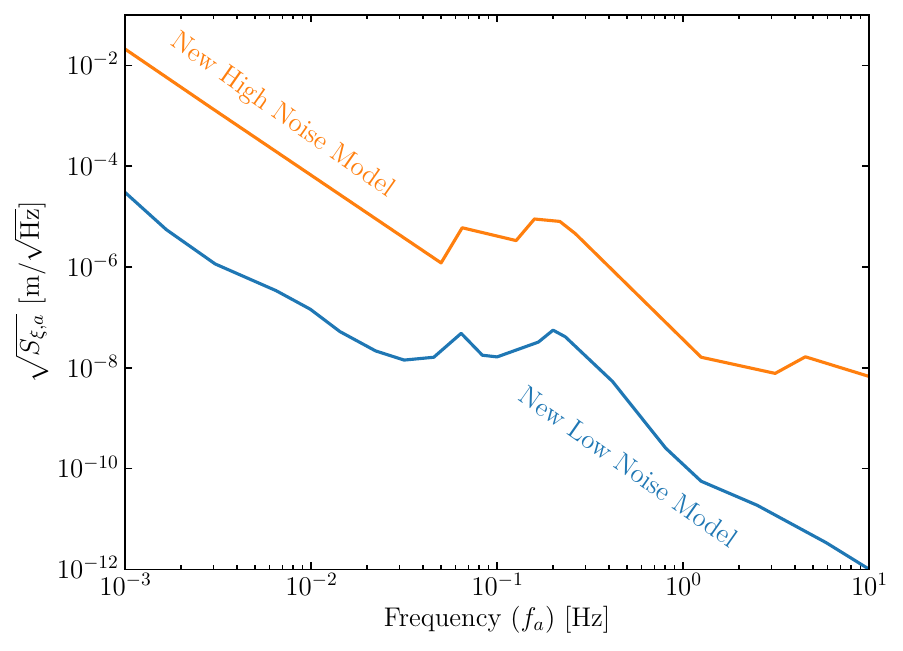}
\caption{Peterson's models of the frequency-dependent spectrum of the vertical displacement $\xi_a$ at the surface of the Earth between \SI{e-3}{Hz} and \SI{10}{Hz}. The orange curve refers to the prediction from the new high noise model (NHNM), while the blue curve is from the new low noise model (NLNM).}
\label{fig: PSD vertical displacement}
\end{figure}

In this work, we employ the homogeneous half space model for the ground around the interferometer. 
This model implies that the ground is isotropic around the detector and that only the fundamental Rayleigh mode is present. 
The fundamental mode travels horizontally at the Earth-air interface and is present irrespective of the geological properties of the site. 
Following the GGN treatments for LIGO~\cite{hughes_seismic_1998,Harms:2015zma}, we represent the seismic field as an incoherent superposition of monochromatic plane-waves propagating isotropically at the Earth's surface. 
The amplitude of each wave depends on the vertical displacement $\xi_a$ at the Earth's surface, which depends on the wave frequency,
and is characterised by an exponentially decaying amplitude while moving away from the surface in either direction.
Here, we bound the upper and lower values of the vertical displacement using Peterson's new high noise model (NHNM) and new low noise model (NLNM)~\cite{USGS}. 
In Fig.~\ref{fig: PSD vertical displacement}, we display the vertical displacement spectrum $\sqrt{S_{\xi, a}}$ according to Peterson's model, where ${S_{\xi, a}} \propto \left \langle \xi_a^2 \right \rangle$ is the power spectral density of the vertical displacement (see section~\ref{sec:AnalysisMultiGrad} and appendices referenced therein for more details).
We limit our analysis to frequencies greater than $10^{-3}$\,Hz since below this frequency the Earth's finite size is expected to significantly alter the properties of the seismic background and render our analysis based on Rayleigh waves invalid~\cite{Nishida, Harms:2015zma,10.2307/j.ctv131bvfd.7}. 

Before moving on to present the results for the induced phase shift, we briefly consider other potential sources that could induce time-dependent perturbations to the gravitational potential and thereby contribute to the gravity gradient noise. Firstly, we do not include the effect from atmospheric density fluctuations, which dominate over the Rayleigh wave-induced contribution below $\sim0.1$\,Hz in horizontal atom interferometer configurations~\cite{MIGAconsortium:2019efk}. 
Secondly, seismic waves could scatter from the shaft's surfaces, but these effects are negligible for cavity radii much smaller than the wavelength of the seismic waves~\cite{Harms:2015zma}. For the frequency range between $10^{-3}$\,Hz and 1\,Hz, the wavelength of these perturbations falls between $\sim\SI{40}{m}$ and $\sim\SI{e4}{m}$, while the expected radial size of the access shaft is $\mathcal{O}(\SI{5}{m})$. Thirdly, we also neglect the direct effect induced by vibrations on the laser system, since we expect that these vibrations can be mitigated with suspensions and other techniques widely used in long-baseline interferometry~\cite{MAGIS-100:2021etm}. Finally, anthropic sources such as heavy vehicles could play a role; owing to their intrinsic site-specific dependence, they will be considered in a separate study.

Returning to the assumption that the seismic field arises from an incoherent superposition of fundamental Rayleigh modes, we obtain the angle-averaged seismically-induced gravitational potential experienced by a test mass underground at time $t$ and position $\boldsymbol{r}_i = (0,0,z_i)$ by integrating over all ground density fluctuations around the experiment (recall that $z_i$ is the vertical position from the top of the interferometer shaft).
In Appendix~\ref{app:RayleighGGN}, we derive the result that the potential due to seismic fluctuations takes the form
\begin{equation}
\begin{split}
V\left(z_i, t\right) \propto & \sum_a \xi_a\, \Bigg [-2 \exp \left(-q \frac{\omega_a z_i}{c_H} \right) \\ & + \left(1+\sqrt{q/s}\right) \exp \left(-\frac{\omega_a z_i}{c_H}\right)  \Bigg ] \, \cos \left( \omega_a t + \widetilde{\theta}_a \right ) \, ,
\end{split}
\label{eq:GGNPotential}
\end{equation}
where $q$ and $s$ are $\mathcal{O}(1)$ dimensionless parameters that depend exclusively on the geological properties of the ground, $c_H$ is the horizontal speed of the Rayleigh wave, $\widetilde{\theta_a}$ is a random phase and the sum is over discrete frequencies $\omega_a$ as in the ULDM case (see Eq.~\eqref{eq:full-field}). With this potential, we find that the GGN phase shift recorded by the $i$\textsuperscript{th} AI at the end of a sequence that starts at time $m\Delta t$ can be expressed as
\begin{equation}\label{eq:GGNPhase}
\begin{split}
    \Phi^{(i)}_{\mathrm{GGN},m} = & \sum_{a}  \xi_a \Bigg [\widetilde{A}_a \exp \left (-q\frac{\omega_a z_i}{c_H} \right) + \widetilde{B}_a \exp \left ( -\frac{\omega_a z_i}{c_H} \right) \Bigg ] \cos{\widetilde{\phi}_{a,m}}\;,
\end{split}   
\end{equation}
 where $\widetilde{A}_a$ and $\widetilde{B}_a$ depend on experimental and geological parameters, and $\widetilde{\phi}_{a,m} \supset \omega_a m\Delta t + \widetilde{\theta}_a$ is the phase of the wave at the end of the sequence. In Appendix~\ref{app:RayleighGGN} we provide a full derivation of this result, together with complete expressions for~$\widetilde{A}_a$ and~$\widetilde{B}_a$. Up to a coefficient,
 \begin{equation}
     \widetilde{A}_a, \widetilde{B}_a \propto \frac{n}{\omega_a^2} \, \,\sin^2 \left(\frac{\omega_a T}{2}\right) \, ,
     \label{eq: GGN amp 1}         
 \end{equation}
which implies that the amplitude of the GGN phase, as for the ULDM phase shift, is a function of~$n$ and~$T$. 

The random variables $\widetilde{\theta}_{a}$ and $\xi_a$ that enter Eq.~\eqref{eq:GGNPhase} are uncorrelated random variables. As for the ULDM case, this means that over long timescales the GGN phase measured by a gradiometer will be characterised by a vanishing expectation value,
 \begin{equation}
\begin{aligned}
\Big \langle \Phi^{(i,j)}_{\mathrm{GGN}, m} \Big \rangle &= 0 \, ,
\end{aligned}
\label{eq:GGNexpectation}
\end{equation}
and, assuming sufficiently similar geological conditions along the baseline, covariance given~by
\begin{equation}
\begin{split}
\Big \langle \Phi^{(i,j)}_{\mathrm{GGN}, m} \Phi^{(i',j')}_{\mathrm{GGN}, m'}\Big \rangle 
& =\sum_{a}  \frac{\langle \xi_a^2 \rangle}{2} F_{\mathrm{GGN},a}^{(i,j)}F_{\mathrm{GGN},a}^{(i',j')} \times \cos (\omega_a \Delta t(m-m'))  \, ,
\end{split}
\label{eq:GGNcovariance}
\end{equation}
where 
\begin{gather} \label{eq: GGN amp}
        F_{\mathrm{GGN},a}^{(i,j)} = F_{\mathrm{GGN},a}^{(i)}-F_{\mathrm{GGN},a}^{(j)} \, , \\ 
        F_{\mathrm{GGN},a}^{(i)} = \widetilde{A}_a \exp \left (-q\frac{\omega_a z_i}{c_H} \right) + \widetilde{B}_a \exp \left ( -\frac{\omega_a z_i}{c_H} \right) \, .
\end{gather}

Unlike ULDM, the GGN phase shift depends on the length scale 
\begin{equation} \label{eq:GGNscalelength}
\lambda_{\rm{GGN}}= \frac{c_H}{\omega_a}\simeq\SI{100}{m}~\left(\frac{250~\mathrm{m\,s^{-1}}}{ c_H}\right)^{-1}~\left(\frac{2.5~\mathrm{Hz}}{\omega_a}\right)\;.
\end{equation}
As will be discussed in detail in section~\ref{sec:ULDMGGNBackground}, one can precisely use $\lambda_\mathrm{GGN}$ to disentangle the ULDM and GGN gradiometer phase shifts. In summary, in the regime where $\lambda_\mathrm{GGN} \gg L$, the GGN (c.f.~Eqs.~\eqref{eq:GGNPhase}-\eqref{eq: GGN amp}) and ULDM (c.f.~Eqs.~\eqref{eq:ULDMphase}-\eqref{eq: ULDM amp}) gradiometer phase shifts are both proportional to the separation between the interferometers, namely $\Delta z$; hence the two phase shifts cannot be distinguished. Instead, in the regime $\lambda_\mathrm{GGN} < L$, the two phase shifts manifestly scale differently with $\Delta z$, which implies that the ULDM and GGN could be disentangled by probing different length scales. 

The values of $c_H$, $s$ and $q$ depend on the precise makeup of the ground where the interferometer is situated. For example, Ref.~\cite{hughes_seismic_1998} quotes a range of $200\,\mathrm{m s}^{-1} \lesssim c_H \lesssim 6000 \,\mathrm{m s}^{-1}$ depending on the rock and mineral composition. 
As we do not have a specific site in mind for this work, we focus on geological parameter values that will most clearly highlight the impact of atom multi-gradiometry and the benefits of the likelihood formalism to potentially extend the sensitivity in the sub-Hz frequency range.
This corresponds to the lower range of $c_H$ where $\lambda_{\rm{GGN}}\lesssim L$ in the frequency ($f$) range  between 0.1\,Hz and 1\,Hz (corresponding to the angular frequency ($\omega$) range between 0.6\,Hz and 6\,Hz). Specifically, our default set of parameters correspond to the surface materials at the LIGO Livingston site~\cite{hughes_seismic_1998}, where the ground density is 1800\,kg\,m$^{-3}$, $c_H=205\,\mathrm{m\, s}^{-1}$, $q=0.88$ and $s=0.36$. In section~\ref{sec:ULDMGGNBackground} we will also briefly consider the scenario where $c_H$ is over $1000\,\mathrm{m\, s}^{-1}$ such that $\lambda_{\rm{GGN}}> L$.

\subsection{Atom shot noise}

The measurement of the phase recorded by an AI is ultimately limited by atom shot noise. Indeed, as the phase shift is extracted from the comparison of atoms in the ground versus excited state at the end of a sequence, the Poissonian statistics governing atom number affect the interferometer phase. 

Formally, the atom shot noise phase of the $i^\mathrm{th}$ interferometer $\Phi_{\mathrm{ASN}}^{(i)}$  is independently sampled at each measurement from a normal distribution with zero mean and variance $\sigma_{\mathrm{Atom}}^2= (C^2 N_\mathrm{Atom})^{-1}$ for phase differences close to $\pi/2$, where $N_\mathrm{Atom}$~is the number of atoms in the cloud and $C \leq 1$ is the interferometer contrast~\cite{Le_Gou_t_2008, PhysRevA.47.3554, Badurina:2021lwr}. For simplicity, in our analysis we set $C=1$.
Because the variance $\sigma_{\mathrm{Atom}}^2$ and mean are frequency-independent, and given that the noise phase is independent in each measurement, the phase contribution from atom shot noise can be modeled as white noise. Hence, the atom shot noise phase statistics can be summarised as 
\begin{align}
\langle \Phi^{(i)}_{\mathrm{ASN}, m}\rangle &= 0 \label{eq:ASNExpectation} \, , \\
\langle \Phi^{(i)}_{\mathrm{ASN}, m}\Phi^{(j)}_{\mathrm{ASN}, m'}\rangle &= \delta_{m m'}\delta^{ij}\sigma_{\mathrm{Atom}}^2 \, .
\label{eq:ASNCovariance}
\end{align}

The atom shot noise contribution to the gradiometer phase will inherit the statistical properties of the phases measured by each interferometer. Recalling that the atom shot noise phase contribution for the ($i,j$)\textsuperscript{th} gradiometer is defined as $\Phi^{(i,j)}_\mathrm{ASN} = \Phi^{(i)}_\mathrm{ASN}-\Phi^{(j)}_\mathrm{ASN}$ and using Eqs.~\eqref{eq:ASNExpectation}-\eqref{eq:ASNCovariance}, in the limit of long integration time, the atom shot noise phase will average to zero,
\begin{align}
\langle \Phi^{(i,j)}_{\mathrm{ASN}, m}\rangle &= 0  \, 
\end{align}
and is characterised by the covariance
\begin{align}
\langle \Phi^{(i,j)}_{\mathrm{ASN}, m}\Phi^{(p,q)}_{\mathrm{ASN}, m'}\rangle &= \delta_{m m'} (\delta^{ip}+\delta^{jq}-\delta^{iq}-\delta^{jp}) \sigma_{\mathrm{Atom}}^2 \, .
\label{eq: atom shot noise covariance}
\end{align}

Unlike the ULDM and GGN phases, Eq.~\eqref{eq: atom shot noise covariance} can be exactly zero when comparing four different interferometers. Hence, in this paper, we expect zero correlations between some gradiometer pairs to be a unique feature of atom shot noise. As discussed in section~\ref{sec:ULDMGGNBackground}, it is precisely this feature that will allow the $\sim\sqrt{\mathcal{N}}$ sensitivity enhancement at high frequencies, where the background is dominated by atom shot noise and $\mathcal{N}$ is the number of interferometers.

\section{Analysis of an atom multigradiometer }\label{sec:AnalysisMultiGrad}

Having characterised the signal and noise correlations of our model, in this section we will present a multigradiometer analysis scheme for DM searches to extract additional information available to two or more gradiometers, which will be of crucial importance to predict and improve an experiment's sensitivity at low frequencies. Our approach makes use of the machinery developed in Refs.~\cite{Foster:2017hbq, Foster:2020fln} for axion-like particle searches. 

Because a DM signal is characterised by a frequency largely set by its mass, and by a frequency spread that is dictated by its speed distribution, we will present all of the subsequent analysis in the frequency domain. Therefore, the appropriate tool for analysing the data is the discrete Fourier transform (DFT) $\widetilde{\Phi}_k^{(i,j)}$ of the phase measured by the $(i,j)$\textsuperscript{th} gradiometer, which is defined as
\begin{equation}
\widetilde{\Phi}_k^{(i,j)} = \sum_{m=0}^{N-1} \Phi_m^{(i,j)} \exp{\left(-\frac{2 \pi \iota k m}{N} \right)}\;,
\label{eq: DFT}
\end{equation}
where $k \in \{0,1,...,N-1\}$. Following Ref.~\cite{Foster:2020fln}, we decompose the DFT in terms of real and imaginary parts, namely
\begin{align}
    R_{k}^{(i,j)}&=\frac{\Delta t}{\sqrt{T_\mathrm{int}}} \operatorname{Re}\left[\widetilde{\Phi}_{k}^{(i, j)}\right] \, , \\ 
    I_{k}^{(i, j)}&=\frac{\Delta t}{\sqrt{T_\mathrm{int}}} \operatorname{Im}\left[\widetilde{\Phi}_{k}^{(i, j)}\right] \, ,
    \label{eq: real and imaginary DFT}
    \end{align}
such that the one-sided power spectral density (PSD) $S_k^{(i,j)}$ of the time-series may be written in terms of this decomposition as
\begin{equation}
    S_k^{(i,j)} \equiv \frac{(\Delta t)^2}{T_\mathrm{int}}\left |\widetilde{\Phi}_{k}^{(i, j)}\right|^2 = \left |R_{k}^{(i,j)}\right|^2 + \left |I_{k}^{(i,j)}\right|^2 \, .
\end{equation}
The information in the $k$-frequency component collected by all gradiometers can then be organized into a  $\mathcal{N}(\mathcal{N}-1)$-dimensional data vector, 
\begin{equation}
    \mathbf{d}_k = \Big [ R_k^{(1,2)}, I_k^{(1,2)},\ldots , R_k^{(\mathcal{N},\,\mathcal{N}-1)}, I_k^{(\mathcal{N},\,\mathcal{N}-1)}\Big]^{\mathrm{\mathbf{T}}} \, .
    \label{eq: data vector}
\end{equation}
Since the phase measured by each interferometer, and hence gradiometer, has an expectation value of zero (cf. Eqs.~\eqref{eq:ULDMexpectation},~\eqref{eq:GGNexpectation}~and~\eqref{eq:ASNExpectation}) and is Gaussian distributed,\footnote{We quote this result from Refs.~\cite{Foster:2017hbq, Foster:2020fln}, in which the authors numerically show that white noise and Rayleigh-distributed data is described by a multi-variable Gaussian distribution.} the data vectors will be characterised by a symmetric $(\mathcal{N}(\mathcal{N}-1) \times \mathcal{N}(\mathcal{N}-1))$-dimensional covariance matrix $\mathbf{\Sigma}_k = \left \langle \mathbf{d}_k\mathbf{d}^\mathrm{\mathbf{T}}_k\right \rangle $ whose entries will depend on both the noise and signal covariances presented in the previous section. Because the noise and signal are uncorrelated, we can use the statistical properties of the different gradiometer phases to decompose the covariance matrix $\boldsymbol{\mathrm{\Sigma}}_{k}$ into signal $\boldsymbol{\mathrm{S}}_{k}$ and noise $\boldsymbol{\mathrm{B}}_{k}$ contributions. In Appendix~\ref{app:CovarianceMatrices} we further elucidate this point and provide detailed analytical calculations for the entries of $\boldsymbol{\mathrm{B}}_{k}$ and $\boldsymbol{\mathrm{S}}_{k}$ given the signal and background models used in this work.

By defining a model $\mathcal{M}$ with parameter vector $\boldsymbol{\theta}$ that has nuisance parameters $\boldsymbol{\theta}_{\mathrm{nuis}}$, which describe GGN and ASN in the individual gradiometers, and signal parameters $\boldsymbol{\theta}_{\mathrm{sig}}$, which characterize the ULDM signal contribution, the likelihood is given by 
\begin{equation}
    \mathcal{L}(d \mid \mathcal{M}, \boldsymbol{\theta})=\prod_{k=1}^{N-1} \frac{\exp \left[-\frac{1}{2} \mathbf{d}_{k}^{T} \cdot \boldsymbol{\Sigma}_{k}^{-1}(\boldsymbol{\theta}) \cdot \mathbf{d}_{k}\right]}{\sqrt{(2 \pi)^{2 \mathcal{N}}\left|\boldsymbol{\Sigma}_{k}(\boldsymbol{\theta})\right|}} \, ,
\end{equation}
where $d = \{\mathbf{d}_k\}$ is the set of data vectors at each $k$-frequency, $|\boldsymbol{\Sigma}_{k}(\boldsymbol{\theta})|$ is the determinant of the covariance matrix and $\boldsymbol{\theta} = \{\boldsymbol{\theta}_{\mathrm{sig}}, \boldsymbol{\theta}_{\mathrm{nuis}}\}$.
In our definition of the likelihood, we omitted the zeroth frequency component, since the DFT of the data at $k = 0$ would be polluted by a tower of static effects, while the ULDM signals considered here are time-dependent. Additionally, the mass of a hypothetical ULDM candidate in this frequency bin would be degenerate with any value below $2\pi/T_\mathrm{int}$, thus leading to poor mass resolution.

With this likelihood, we define the following frequentist tool based on the log profile likelihood
\begin{equation}
\begin{split}
\Theta\left(\boldsymbol{\theta}_{\mathrm{sig}}\right) &= 2\Big[\ln \mathcal{L}\left(d \mid \mathcal{M},\left\{\hat{\boldsymbol{\theta}}_{\text {nuis}}, \boldsymbol{\theta}_{\mathrm{sig}}\right\}\right) \\ &-\ln \mathcal{L}\left(d \mid \mathcal{M},\left\{\hat{\hat{\boldsymbol{\theta}}}_{\text {nuis}}, \boldsymbol{\theta}_{\mathrm{sig}}=\mathbf{0}\right\}\right)\Big] \, ,
\end{split}
\label{eq: log profile likelihood}
\end{equation}
Here, $\hat{\boldsymbol{\theta}}_{\text {nuis}}$ denotes the values of the nuisance parameters that maximise the likelihood in the signal plus background hypothesis, while $\hat{\hat{\boldsymbol{\theta}}}_{\text {nuis}}$ represents the values of the nuisance parameters that maximise the likelihood in the background-only hypothesis.

To set upper limits on couplings to scalar ULDM, we set $\boldsymbol{\theta}_{\mathrm{sig}} = \left (m_{\phi}, d_\phi\right)$, where we remind the reader that $m_{\phi}$ is the dark matter mass and $d_\phi$ is the coupling strength of the relevant linear interaction between ULDM and Standard Model (SM) operators. Then, Eq.~\eqref{eq: log profile likelihood} can be used to define a test statistic for setting upper limits on the ULDM-SM couplings $d_\phi$,
\begin{equation}
    q\left(m_{\phi}, d_\phi\right)= \begin{cases} \Theta\left(\{m_{\phi}, d_\phi\right\})-\Theta\left(\{m_{\phi}, \hat{d_\phi}\}\right) & d_\phi \geq \hat{d_\phi} \\ 0 & d_\phi<\hat{d_\phi}\end{cases}
\end{equation}
where $\hat{d_\phi}$ is the value of $d_\phi$ that maximises $\Theta\left(m_\phi, d_\phi\right)$ at fixed $m_\phi$.\footnote{In light of Eq.~\eqref{eq: log profile likelihood}, this test statistic is independent of the background-only hypothesis, which is in agreement with the literature~\cite{Cowan:2010js}.} With this definition, the test statistic $q$ at fixed $m_\phi$ is described by a half chi-squared distribution with one degree of freedom~\cite{Foster:2017hbq, Cowan:2010js} such that, for a given $m_\phi$, the 95\% limit on $d_\phi$ is set when $q(m_\phi, d_{\phi, 95\%}) \approx -2.70$ when $T_\mathrm{int} \gg \tau_c$ and $q(m_\phi, d_{\phi, 95\%}) \approx -7.55$ when $T_\mathrm{int} \ll \tau_c$, where $\tau_c$ is the coherence time of the ULDM wave~\cite{Centers:2019dyn}.\footnote{To make contact with the signal-to-noise ratio (SNR) defined in Refs.~\cite{Badurina:2019hst, Badurina:2021lwr, Badurina:2021rgt}, please note that $\mathrm{SNR} \sim \sqrt{-q(m_\phi, d_{\phi})}$.} In subsequent sections, for our choice of integration time, the former regime applies for DM masses above $\sim 10^{-17}$~eV - i.e. frequencies above $10^{-2}$~Hz. We have not investigated the precise value of $q$ for setting 95\% exclusion limits in the regime where $T_\mathrm{int}\sim\tau_c$, which is beyond the scope of this work; here, we use the approximation $q(m_\phi, d_{\phi, 95\%}) \approx -2.70$ when $T_\mathrm{int} > \tau_c$ and $q(m_\phi, d_{\phi, 95\%}) \approx -7.55$ when $T_\mathrm{int} \leq \tau_c$.

\subsection{Test statistics in the Asimov approach}\label{subsec:TSAsimov}

To infer the expected distribution of the test statistics and  understand the expected sensitivity of the multigradiometer experiment, many realisations of the experiment can be generated via Monte Carlo simulations. Alternatively, it is also possible to infer the asymptotic properties of the test statistics analytically by taking the dataset to be equal to the mean predictions of the model under consideration and neglecting statistical fluctuations, which is known as the Asimov approach~\cite{Foster:2017hbq, Foster:2020fln}. Since the test statistic for setting upper limits depends on the modified log profile likelihood, we define the basic frequentist tool in the Asimov approach as
\begin{equation}
    \widetilde{\Theta} (\boldsymbol{\theta}_{\mathrm{sig}}) = \langle \Theta ( \boldsymbol{\theta}_{\mathrm{sig}}) \rangle \, .
    \label{eq: Asimov Theta}
\end{equation}
Recalling that we have assumed a covariance matrix $\boldsymbol{\mathrm{\Sigma}}_{k}$ that can be split into signal $\boldsymbol{\mathrm{S}}_{k}$ and noise $\boldsymbol{\mathrm{B}}_{k}$ contributions, and assuming that the signal is parametrically weaker than the background contributions, we may express the Asimov test statistic for setting upper limits as:
\begin{equation}
    \widetilde{q}\left(m_{\phi}, d_\phi\right)= \begin{cases} -\frac{1}{2}\sum_{k=1}^{N-1} \mathrm{Tr}\left [\mathbf{S}_k \mathbf{B}^{-1}_k \mathbf{S}_k \mathbf{B}^{-1}_k \right]  & d_\phi \geq 0 \\ 0 & d_\phi<0\end{cases} \, ,
    \label{eq:UpperLimitAsimov}
\end{equation}
which scales with  $T_\mathrm{int}^2$ when $T_\mathrm{int} \gg \tau_c$, and with $\tau_c \, T_\mathrm{int}$ when $T_\mathrm{int} \ll \tau_c$. In Appendix~\ref{app:AsimovTestStatistics} we provide a derivation of Eq.~\eqref{eq:UpperLimitAsimov}, we explain the origin of the behaviour of the test statistic in the two aforementioned regimes, and we show examples of the expected power spectral density for different values of $T_\mathrm{int}$. Since the non-vanishing entries of $\mathbf{S}_k$ are proportional to $d_\phi^2$, while those of $\mathbf{B}_k$, if assuming atom shot noise only, are proportional to $1/N_\mathrm{atom}$, $\widetilde{q}$ is proportional to $d_\phi^4 \, N_\mathrm{atom}^2$. This is in agreement with results in Ref.~\cite{Badurina:2021rgt} in which $d_\phi \propto 1/\sqrt{N_\mathrm{atom}}$.

\section{ULDM searches in a GGN background}\label{sec:ULDMGGNBackground}

\begin{table}
\begin{center}
\begin{tabular}{ c | c c  c c c c c}
Design & $L$[m] & $T$[s] & $n$ & $\Delta z_\mathrm{max}$[m] & $N_\mathrm{atom}$ & $\Delta t$[s] & $T_\mathrm{int}$[s] \\
\hline
Intermediate & $100$ & 1.4 & $1000$ & 85 & $10^8$ & 1.5 & $10^8$ \\
Advanced & $1000$ & 1.7 & $2500$ & $970$ & $10^{10}$ & 1.0 & $10^8$ \\
\end{tabular}
\end{center}
\caption{The experimental parameters for the different atom interferometer scenarios considered in this work. The `intermediate' and `advanced' designs could be implemented in versions of future vertical gradiometers, such as AION-100 and MAGIS-100, and AION-km, respectively. $\Delta z_\mathrm{max}$ refers to the maximum gradiometer length given the choice of interferometer parameters. We also consider scenarios where $\Delta z$ is shorter than the maximum value.} 
\label{table:ExperimentalParameters}
\end{table}

In order to assess the impact of GGN on a gradiometer's maximum sensitivity, it is of paramount importance to understand the relative size of GGN with respect to atom shot noise. Indeed, it will be precisely this comparison that will guide our analysis into understanding the benefits of a multigradiometer experiment to mitigate and, in an ideal scenario, entirely cancel GGN at particular frequencies. Additionally, because the GGN background as detected by an atom interferometer is largely dependent on experimentally tunable parameters, we will provide a discussion in terms of two vertical gradiometer configurations: an `intermediate' and an `advanced' design, which feature a \SI{100}{m} and \SI{1}{km} baseline, respectively. In Table~\ref{table:ExperimentalParameters} we summarise all of the relevant parameters that will be necessary for subsequent calculations and discussions. Additionally, we will restrict our discussion to the impact of GGN on the sensitivity to scalar ULDM couplings to the electron mass, i.e.\ $d_\phi = d_{m_e}$. We will begin our discussion with a single atom gradiometer before considering the advantages brought by an atom multigradiometer in mitigating the GGN background.

\begin{figure*}[t!]
    \includegraphics[width=.5\columnwidth]{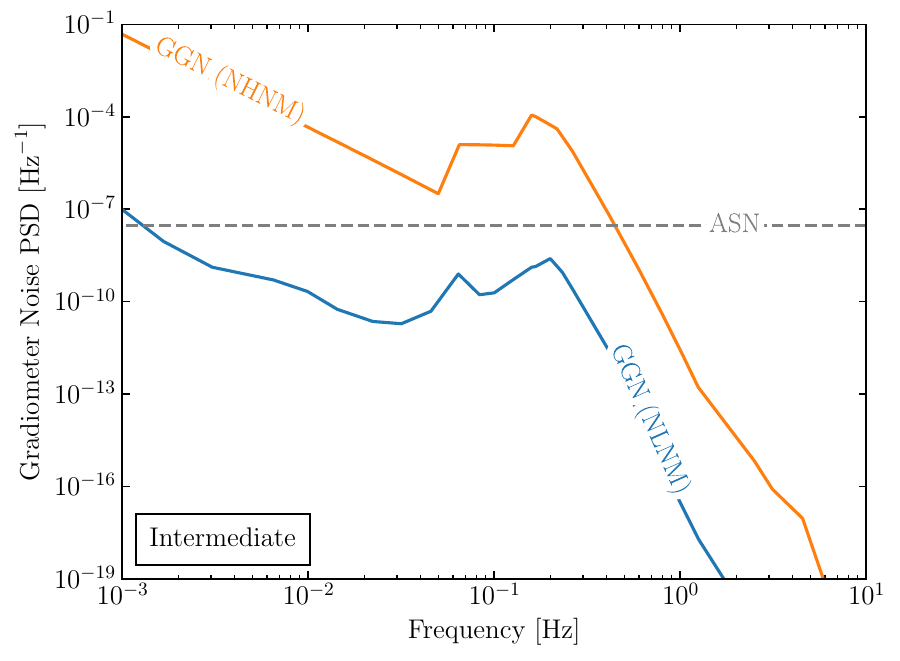} 
\includegraphics[width=.495\columnwidth]{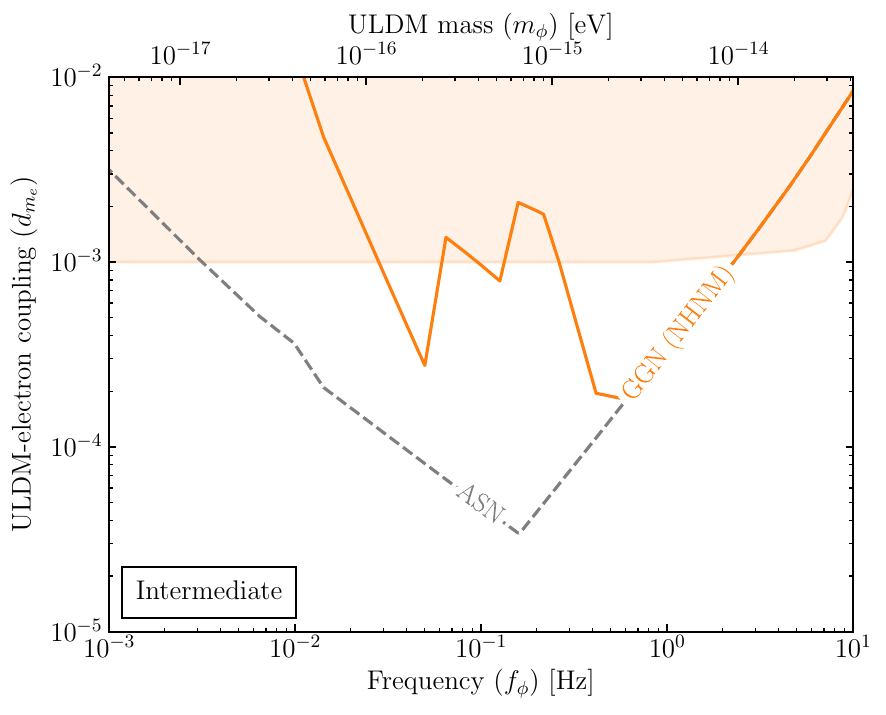}
    \caption{  Impact of GGN on a single atom gradiometer ($\mathcal{N}=2$) for the `intermediate' design defined in Table~\ref{table:ExperimentalParameters}.
 Left panel: Contributions to the gradiometer noise PSD as a function of frequency from atom shot noise (dashed grey) and GGN according to the NHNM (orange) and NLNM (blue) scenarios. Right panel: Projected 95\% CL exclusion curves on the ULDM-electron coupling when assuming an atom shot noise-only background (grey dashed), and when GGN modelled by the NHNM is also included (orange). The NLNM result is degenerate with the ASN line for frequencies greater than $\approx 10^{-3}$~Hz, so is omitted for clarity. The orange shaded region is excluded by MICROSCOPE~\cite{PhysRevLett.129.121102}. In high seismic-noise scenarios, GGN can limit an `intermediate' design's reach in ULDM parameter space. 
}
\label{fig:SensitivityAION100_N2}
\end{figure*}

\subsection{Single atom gradiometer}\label{sec:ULDMGGNBackgroundSingleGradiometer}

In the left panel of Fig.~\ref{fig:SensitivityAION100_N2} we plot the single atom gradiometer ($\mathcal{N}=2$) noise power spectral density (PSD) as a function of frequency for the `intermediate' design defined in Table~\ref{table:ExperimentalParameters}.
The atom shot noise (ASN) contribution is shown with a dashed grey line; the gravity gradient noise (GGN) contributions according to the new high (NHNM) and low noise models (NLNM) are shown with orange and blue solid lines, respectively. We see that the PSD of the gradiometer phase induced by GGN dominates the low frequency range below $\sim 0.5$~Hz for the NHNM. This indicates that in an environment where the seismic noise is high, GGN can limit an experiment's reach in ULDM parameter space for $L\sim \mathcal{O}{(\SI{100}{m})}$.\footnote{In the frequency range of interest, GGN does not limit the sensitivity of $L\sim\SI{10}{m}$ baselines, even in the NHNM scenario.}

We demonstrate the impact of the NHNM model explicitly in the right panel of Fig.~\ref{fig:SensitivityAION100_N2}, which shows the sensitivity projections in the ULDM frequency (or mass) versus ULDM-electron coupling plane for the `intermediate' design. Typically, the sensitivity projections oscillate as a function of the ULDM mass. However, for clarity, we follow Refs.~\cite{Arvanitaki:2016fyj, Badurina:2021lwr} and plot only the envelope of the oscillations by employing the approximation $|\sin x\,|=\mathrm{min}\{x,1/\sqrt{2}\}$ in the phase formulas (e.g., Eqs.~\eqref{eq:app:ULDMphase} and~\eqref{eq:app:GGN}). The dashed grey line shows the projected 95\%~CL exclusion curve in the presence of a background that only contains atom shot noise. In contrast, the solid orange line shows the 95\%~CL exclusion curve when GGN modelled by the NHNM is included. As anticipated from the left panel, the sensitivity projections are reduced below $\sim 0.5$~Hz. For clarity the right panel does not show the projection for the NLNM scenario. As may be anticipated from the left panel, the sensitivity projection for the NLNM is degenerate with the grey line as it is sub-dominant to atom shot noise for frequencies greater than $\approx 10^{-3}$~Hz.

We highlight that the shape of the ASN-limited sensitivity curve can be understood by carefully studying the amplitude of the ULDM gradiometer phase shift (c.f. Eqs.~\eqref{eq:ULDMphase}-\eqref{eq: ULDM amp},~\eqref{eq:app:ULDM amp}). The amplitude is maximised at $m_\phi\sim\pi/T$, or equivalently $f_\phi \sim 1/2T$. In the regime where $m_\phi \ll \pi/T$, the ULDM signal amplitude is proportional to $m_\phi$. Hence, in the limit where $m_\phi \rightarrow 0$, the ULDM-induced phase shift vanishes, so the sensitivity curve rises to large values of $d_\phi$. In the regime where $m_\phi \gg \pi/T$ the ULDM signal amplitude is proportional to $1/m_\phi^2$.  Hence, in the limit $m_\phi \rightarrow \infty$, the ULDM-induced phase shift also vanishes, and the sensitivity curve rises to large values of $d_\phi$. The discontinuity at $\sim 10^{-2}$~Hz is due to the abrupt change in the value of $q$ for setting upper limits (c.f. section~\ref{sec:AnalysisMultiGrad}).

\begin{figure*}
    \includegraphics[width=.5\columnwidth]{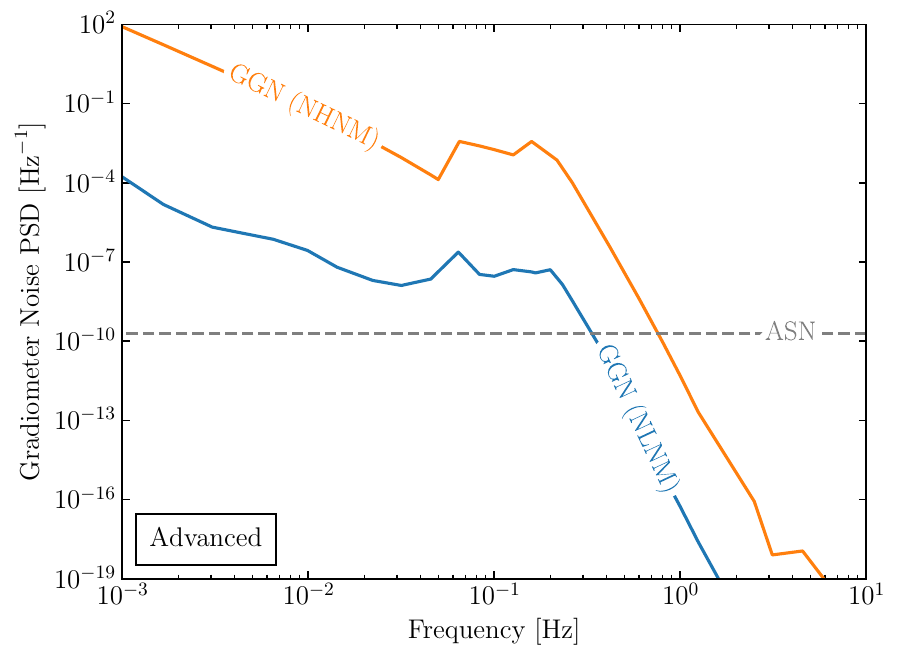} 
    \includegraphics[width=.495\columnwidth]{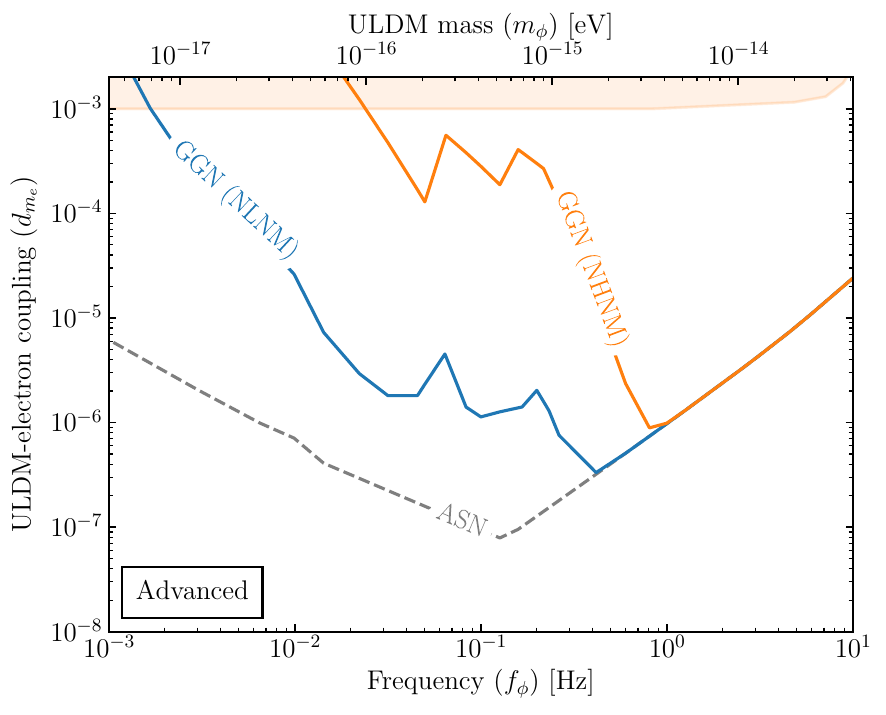}
\caption{
  Impact of GGN on a single atom gradiometer ($\mathcal{N}=2$) for the `advanced' design defined in Table~\ref{table:ExperimentalParameters}.
  Left panel: Contributions to the gradiometer noise PSD as a function of frequency from atom shot noise (dashed grey) and GGN according to the NHNM (orange) and NLNM (blue) scenarios. Right panel: Projected 95\% CL exclusion curve on the ULDM-electron coupling with an atom shot noise-only background (grey), and when GGN according NHNM (orange) and NLNM (blue) are also included. The orange shaded region is excluded by MICROSCOPE~\cite{PhysRevLett.129.121102}. GGN can greatly diminish an `advanced' design's reach in ULDM parameter space. } 
\label{fig:SensitivityAION1km_N2}
\end{figure*}

Next we turn to the `advanced' design, in which the baseline is 1\,km. The left panel of Fig.~\ref{fig:SensitivityAION1km_N2} again shows the noise PSD for a single atom gradiometer.
For this design, we see that GGN exceeds atom shot noise in the sub-Hz range. In the NHNM (orange) it is the dominant noise source below $\sim 1$~Hz while in the NLNM (blue), differently from the `intermediate' design, it dominates below $\sim 0.5$~Hz as a result of different experimental parameters. Indeed, the PSD of ASN is inversely proportional to the number of atoms, while the PSD of GGN, in the low frequency limit, is roughly proportional to the product $n^2 L^2 T^4$. Since the `advanced' design features twice as many LMT kicks, comparable interrogation and sampling time, a ten-fold increase in the length of the baseline and a one-hundred-fold increase in the number of atoms with respect to the `intermediate' design, the `advanced' design would be characterised by a larger GGN at low frequencies and smaller ASN across the frequency range of interest. 

The right panel of Fig.~\ref{fig:SensitivityAION1km_N2} shows the impact of GGN in the ULDM mass versus ULDM-electron coupling plane. In the sub-Hz regime, in the NLNM scenario the 95\% CL exclusion curve (blue) is weaker by approximately an order of magnitude with respect to a naive estimate accounting for atom shot noise only. The loss of sensitivity in the NHNM is much more dramatic, with a suppression by up to four orders of magnitude with respect to the atom shot noise only result. Hence, we conclude that the sensitivity of a single atom gradiometer ($\mathcal{N}=2$) of baseline length $L\gtrsim \mathcal{O}{(\SI{1}{km})}$ would be greatly diminished across the entire low frequency range in both the NHNM and NLNM.
Nevertheless, in light of the ULDM signal's scaling with $L$ and $n$, the `advanced' design would still be able to probe larger regions of scalar ULDM parameter space (c.f. the right panel of Fig.~\ref{fig:SensitivityAION1km_N2}) than the `intermediate' design (c.f. the right panel of Fig.~\ref{fig:SensitivityAION100_N2}).

Even within the limits of a single atom gradiometer ($\mathcal{N}=2$), there is scope to regain parts of parameter space by tuning the experimental parameters to minimise the impact of GGN.
We observe that the GGN and the ULDM signal in an atom interferometer both depend identically on the number of LMT kicks $n$ and interrogation time $T$ (cf.\ Eqs.~\eqref{eq: ULDM amp} and~\eqref{eq: GGN amp}), but show different scaling relationships with respect to the vertical position along the baseline. Hence, we will focus on the impact of changing the position of the interferometers along the baseline. 

\begin{figure*}
    \centering
    \includegraphics[width=.495\textwidth]{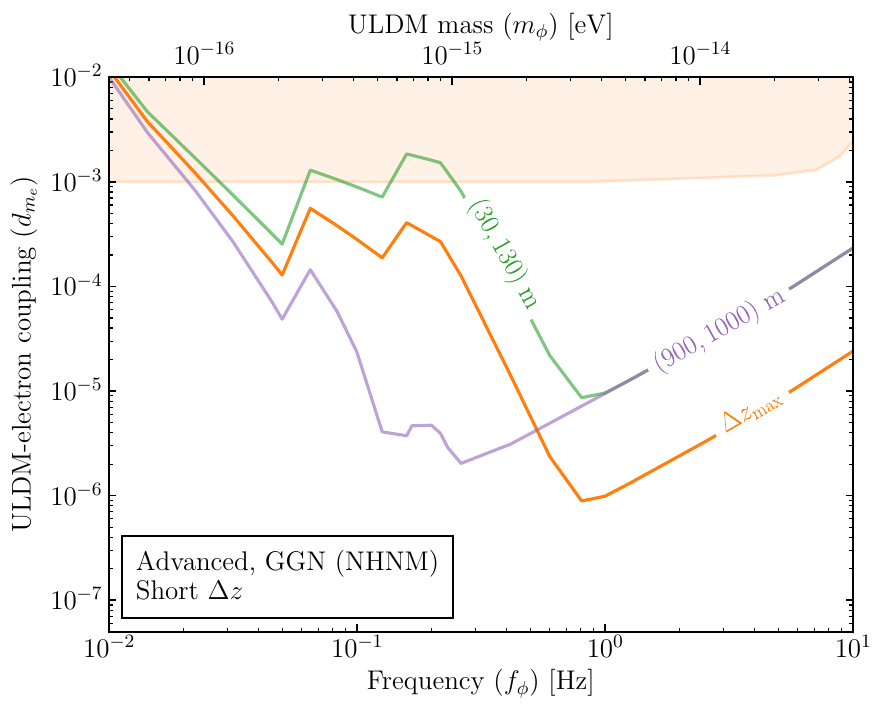} 
    \includegraphics[width=.495\textwidth]{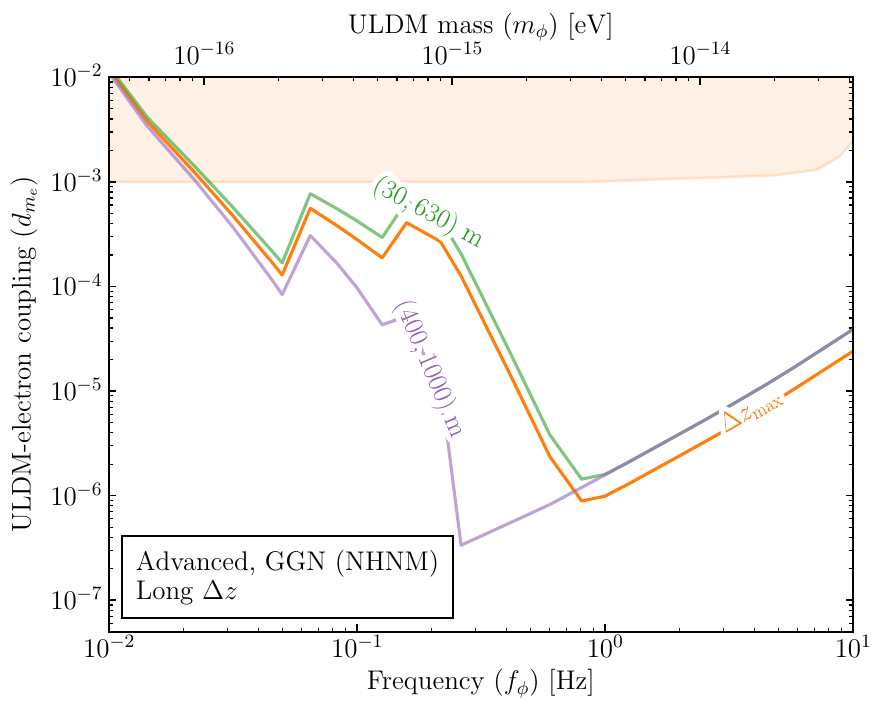}
    \caption{Impact on the 95\% CL exclusion curves for different values of $\Delta z$ and different atom interferometer positions for a single atom gradiometer ($\mathcal{N}=2$) assuming that GGN is modelled by the NHNM.
The orange lines in both panels show the `advanced' design parameters configuration. We also show exclusion curves with interferometers located towards the Earth's surface (green) and towards the bottom of the shaft (purple) assuming $\Delta z = 100$\,m (left panel) and $\Delta z = 600$\,m (right panel) and with all other `advanced' parameters unchanged. The orange shaded region is excluded by MICROSCOPE~\cite{PhysRevLett.129.121102}.
Placing atom interferometers towards the bottom of the shaft regains parts of parameter space below $\sim 0.5$\,Hz but at the cost of reduced sensitivity at higher frequencies.}
    \label{fig:DiffDeltaz}
\end{figure*}

Both panels in Fig.~\ref{fig:DiffDeltaz} show the 95\% CL exclusion curves for the `advanced' design; however, in these panels, we exploit the freedom to locate the two atom interferometers at different positions along the 1\,km baseline.
In both panels the orange line repeats the result from Fig.~\ref{fig:SensitivityAION1km_N2} for the NHNM GGN scenario in which the distance $\Delta z$ between the two interferometers is set to the maximum value, $\Delta z_{\rm{max}}=970$\,m. The left panel of Fig.~\ref{fig:DiffDeltaz} additionally shows the 95\% CL exclusion curves when the atom interferometers are separated by $\Delta z=100$\,m and are situated closer to the Earth's surface (green) or towards the bottom of the shaft (purple). While situating the atom interferometers towards the Earth's surface brings no benefit relative to $\Delta z_{\rm{max}}$, for both values of $\Delta z$ we observe a sensitivity improvement for the configuration at the bottom of the shaft below $\sim 0.5$\,Hz. However, this comes at the cost of losing sensitivity at frequencies greater than $\sim 0.5$\,Hz. The right panel of Fig.~\ref{fig:DiffDeltaz} considers the scenario where $\Delta z=600$\,m,
and the green and purple lines again show the exclusion curves when positioning one of the interferometers closer to the Earth's surface and towards the bottom of the shaft, respectively. In this case, a similar behaviour is observed: the configuration with an AI at the bottom of the shaft is favoured below $\sim 0.7$\,Hz but again at the cost of reduced sensitivity at higher frequencies. Comparing the left and right panels of Fig.~\ref{fig:DiffDeltaz} we see that in a narrow frequency range between $\sim 0.2$\,Hz and $\sim 0.7$\,Hz,
there is a preference for larger $\Delta z$ values at the bottom of the shaft.

\begin{figure}
    \centering
    \includegraphics[width=.495\textwidth]{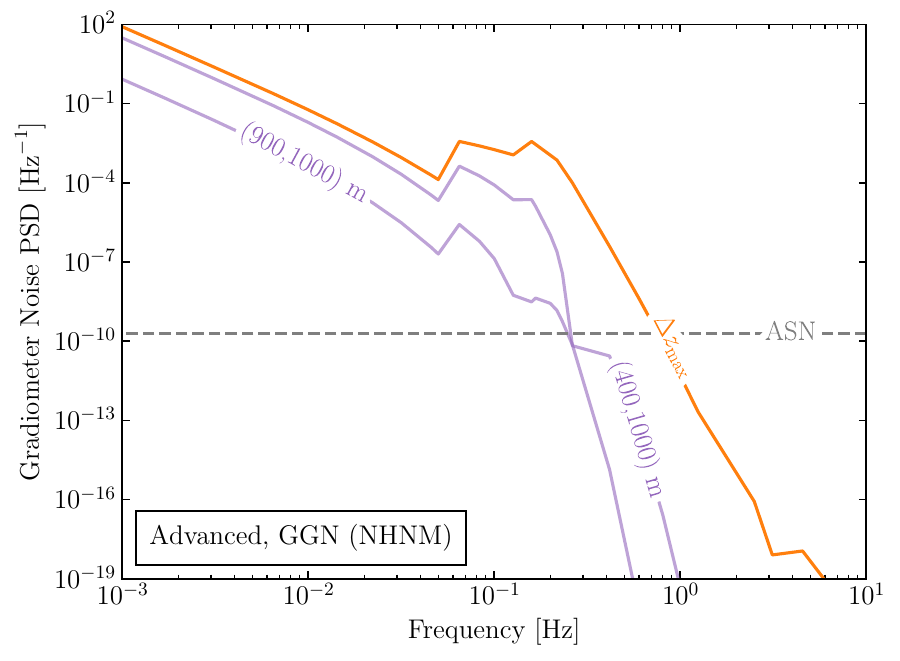} 
    \caption{Contributions to the gradiometer noise PSD from atom shot noise (dashed grey) and GGN according to the NHNM for an `advanced' single atom gradiometer ($\mathcal{N}=2$) in which the AIs are positioned at: 900\,m and 1000\,m ($\Delta z=100$\,m); and 400\,m and 1000\,m ($\Delta z=600$\,m). We also show the PSD of GGN according to the NHNM for $\Delta z_\mathrm{max}=970$\,m, which is identical to orange curve in the left panel of Fig.~\ref{fig:SensitivityAION1km_N2}.}
    \label{fig:PSDDeltaz}
\end{figure}

The frequency-dependent behaviour observed in Fig.~\ref{fig:DiffDeltaz} can be qualitatively understood by considering the dependence of the GGN and ULDM phase shifts on the interferometer position along the baseline with respect to the Rayleigh wave's profile underground.
At fixed $\Delta z$, the GGN background and ULDM signal scale differently in the high and low frequency limit. At low frequencies, corresponding to $\omega_a \ll 1/T$ and $\lambda_{\rm{GGN}} \gg \Delta z$, the GGN and ULDM phase shifts both scale like $\omega_a \Delta z$ (cf.\ Eqs.~\eqref{eq:GGNPhase} and~\eqref{eq:ULDMphase}). 
Therefore, given the geological parameters used in this work, little to no sensitivity enhancement is gained from probing different gradiometer length scales for angular frequencies $\omega_a \ll 10^{-1}$~Hz, where all curves asymptote towards the same value.

In contrast, in the frequency window between $\sim 0.05$\,Hz and $\sim 0.2$\,Hz, where GGN dominates the background, and when $\lambda_{\rm{GGN}} \lesssim \Delta z_\mathrm{max}$, the ULDM and GGN phase shifts scale very differently: as $\Delta z/\omega_a$ for ULDM;  and as $|\exp (-\, \omega_a z_i/c_H) - \exp(-\omega_a z_j/c_H)|/\omega_a^2$ for GGN.
Writing $z_j = z_i + \Delta z$, the ratio between the amplitudes of the ULDM and GGN phase shifts is maximised for $z_i = L$ and $\Delta z$ small, which implies that the GGN background is better mitigated when placing the interferometers at the bottom of the baseline and choosing small gradiometer separations. At high frequencies where the background is dominated by atom shot noise, which is frequency-independent, the maximum reach can be achieved by choosing the largest possible gradiometer length, $\Delta z_\mathrm{max}$. 

Finally, the preference for larger $\Delta z < \Delta z_\mathrm{max}$ values at the bottom of the shaft between $\sim 0.2$\,Hz and $\sim 0.7$\,Hz arises as a result of the fact that at these frequencies there exist positions along the baseline where GGN is subdominant. This can be seen in Fig.~\ref{fig:PSDDeltaz}, where we show the PSD of GGN over frequencies for long and short values of $\Delta z < \Delta z_\mathrm{max}$ when one of the interferometers is located at 1000\,m below the Earth's surface. For both $\Delta z= 600$~m and $\Delta z= 100$~m, the experiment will be ASN-limited above $\sim 0.2$~Hz, which implies that the experimental sensitivity will scale like $\Delta z$ above $\sim 0.2$~Hz. Hence, between $\sim 0.4$~Hz and $\sim 1$~Hz, the configuration with $\Delta z= 600$~m is favoured, as shown in Fig.~\ref{fig:DiffDeltaz}.

\subsection{An atom multigradiometer experiment} \label{sec:ULDMGGNBackgroundMultiGradiometer}

In our discussion of a single atom gradiometer, we observed that GGN can greatly diminish the reach in ULDM parameter space. 
Furthermore, we observed that the impact of GGN could be changed by varying the position of the atom interferometers and the distance $\Delta z$ between them. 
For instance, the configuration with $\Delta z=100$\,m where the atom interferometers are positioned towards the bottom of the 1\,km shaft improved the sensitivity in the sub-Hz regime but at the cost of a loss in sensitivity above 1\,Hz compared to the $\Delta z=970$\,m configuration.
With these observations in mind, we can immediately see the benefit of considering a multigradiometer experiment since, by positioning multiple interferometers along the baseline, it will be possible
to probe the long and short $\Delta z$ configurations \textit{simultaneously} and thereby maximise the sensitivity across all frequencies.

Although $\mathcal{N}(\mathcal{N}-1)/2$ unique gradiometer  measurements can be performed on $\mathcal{N}$ interferometers, at most $\mathcal{N}-1$ gradiometers can be combined in the multigradiometer likelihood-based analysis presented in section~\ref{sec:AnalysisMultiGrad}. This follows from the non-singularity condition of the data's covariance matrix $\boldsymbol{\mathrm{\Sigma}}_{k}$ and can be heuristically understood as the consequence of requiring that gradiometer measurements in a given set are linearly independent, i.e.\ that no gradiometer measurement be written in terms of the sum of other gradiometer measurements. For example, for three AIs, three gradiometer measurements can be performed, namely $\{ \Phi^{(1,2)}, \Phi^{(1,3)}, \Phi^{(2,3)}\}$; since $\Phi^{(i,j)} = \Phi^{(i,k)} - \Phi^{(j,k)}$ for $i,j,k \in \{1, \ldots \mathcal{N}\}$, a multigradiometer analysis containing measurement $\Phi^{(1,2)}$ cannot contain both $\Phi^{(2,3)}$ and $\Phi^{(1,3)}$, but can contain $\Phi^{(1,2)}$ and either $\Phi^{(2,3)}$ or $\Phi^{(1,3)}$, so two gradiometer measurements in total. Alternatively, this statement can be recast in the language of graph theory. Mapping an AI to a vertex and a gradiometer measurement to an edge, a set of gradiometer measurements can be understood as a graph in which two vertices can be connected by, at most, one edge and no cycles are present; the former follows from the requirement that measurements must not be degenerate, while the latter follows from the requirement that gradiometer measurements in a set are linearly independent. Recalling that gradiometer measurements are insensitive to the ordering of interferometer pairs (i.e.\ are undirected), the graphs above are defined as trees (i.e.\ connected undirected graphs that contains no cycles). By definition, it then follows that a tree on $\mathcal{N}$ vertices contains $\mathcal{N}-1$ edges~\cite{wilson10}, so a set of gradiometer measurements on $\mathcal{N}$ AIs contains $\mathcal{N}-1$ unique and linearly-independent elements. 

After fixing the interferometers' positions along the baseline, the number of sets of cardinality $\mathcal{N}-1$ containing unique and linearly-independent gradiometer measurements can be efficiently calculated in graph theory. Indeed, via Caylely's formula, the number of trees on $\mathcal{N}$ labeled vertices is $\mathcal{N}^{\mathcal{N}-2}$~\cite{wilson10}; hence, the number of sets each containing $\mathcal{N}-1$ gradiometer measurements that can be combined in a multigradiometer likelihood analysis is $\mathcal{N}^{\mathcal{N}-2}$.

\begin{figure*}[t!]
    \centering
    \includegraphics[width=0.8\textwidth]{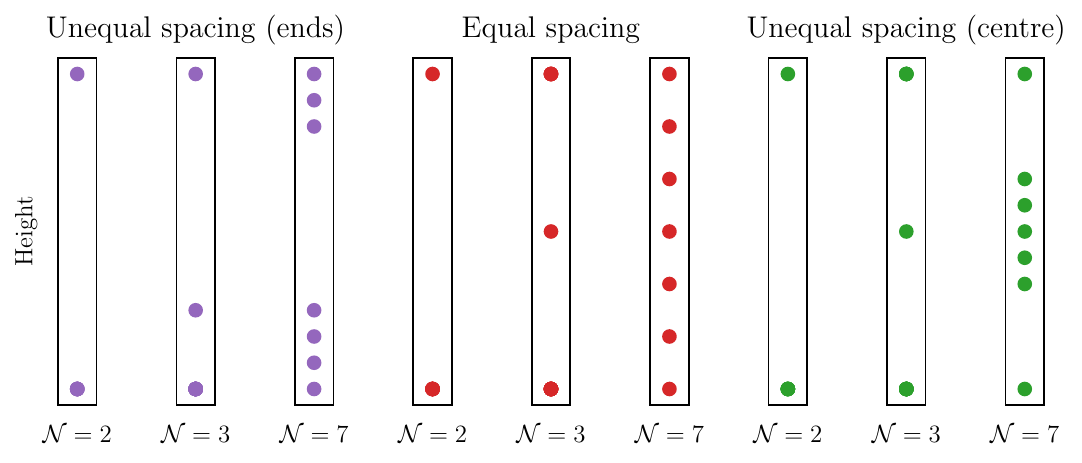}
    \caption{Schematic representation of different interferometer configurations for different interferometer numbers: $\mathcal{N}=2$, $\mathcal{N}=3$ and $\mathcal{N}=7$. With purple dots we show the positioning of interferometers in `unequal spacing (ends)' configurations. We show in red the positioning of interferometers in `equal spacing' configurations. In green we show the `unequal spacing (centre)' configurations.}
    \label{fig:configurations}
\end{figure*}

Any choice of $\mathcal{N}-1$ gradiometer pairs out of $\mathcal{N}^{\mathcal{N}-2}$ possibilities gives the same sensitivity to ULDM. Indeed, for any choice of $\mathcal{N}-1$ gradiometers there exists a map, which preserves the value of the test statistic, from the corresponding data vector to that of any other choice of $\mathcal{N}-1$ interferometer pair. For example, let $\mathcal{N}=3$; the set of possible gradiometer measurements is given by $\{ \Phi^{(1,2)}, \Phi^{(1,3)}, \Phi^{(2,3)}\}$. Suppose that we choose $\{ \Phi^{(1,2)}, \Phi^{(1,3)}\}$ in our likelihood; then, using Eq.~\eqref{eq: real and imaginary DFT}, the corresponding data vector is $\mathbf{d}_k = (R^{(1,2)}, I^{(1,2)}, R^{(1,3)}, I^{(1,3)} )^\mathrm{T} \propto (\Phi^{(1,2)}, \Phi^{(1,2)}, \Phi^{(1,3)}, \Phi^{(1,3)} )^\mathrm{T}$. Now suppose that we want to perform the multigradiometer analysis using a different set of measurements, namely $\{ \Phi^{(2,3)}, \Phi^{(1,3)}\}$, whose corresponding data vector is $\mathbf{d}'_k = (R^{(2,3)}, I^{(2,3)}, R^{(1,3)}, I^{(1,3)} )^\mathrm{T} \propto (\Phi^{(2,3)}, \Phi^{(2,3)}, \Phi^{(1,3)}, \Phi^{(1,3)} )^\mathrm{T}$. It can be shown that there exists a matrix $\mathbf{M}$ which maps $\mathbf{d}_k$ to $\mathbf{d}'_k$ and effectively corresponds to the index relabeling $1\rightarrow3,3\rightarrow1$ and $2\rightarrow2$. In geometric terms, this can be understood as a relabelling of vertices in a graph.  Recalling that the data's covariance matrix $\mathbf{\Sigma}_k = \left \langle \mathbf{d}_k\mathbf{d}^\mathrm{\mathbf{T}}_k\right \rangle$ can be decomposed into signal and noise contributions as $\mathbf{\Sigma}_k = \mathbf{S}_k + \mathbf{B}_k$, it follows that $\mathbf{S}'_k = \mathbf{M}\mathbf{S}_k \mathbf{M}^T$ and $\mathbf{B}'_k = \mathbf{M}\mathbf{B}_k \mathbf{M}^T$. Using the cyclicity of the trace and Eq.~\eqref{eq:UpperLimitAsimov}, the test statistics for the two different choices of gradiometer measurements are equal at all frequencies, i.e.\ $q'_k = q_k$ for all $k$ in the spectrum. Hence, an atom multigradiometer experiment is insensitive to the choice of interferometer pairs.

Having identified the maximum number of gradiometer measurements that can be combined in a multigradiometer analysis and after elucidating the equivalence in the choice of gradiometer measurements, we are in a position to determine which configurations maximise the experimental reach across the frequency band of interest. The optimal choice of configuration, however, is noise model dependent and so warrants a careful analysis. 
To gain intuition, we will explore the effect of different interferometer layouts on the reach of multigradiometer experiments in baselines of length $L\sim \mathcal{O}(\SI{1}{km})$.

\subsubsection{Example configurations}

 To facilitate the comparison between different interferometer configurations, we will focus our analysis on three distinct multigradiometer layouts: unequally-spaced intereferometers clustered at either end of the baseline; equally-spaced configurations; and unequally-spaced configurations in which each end of the baseline is equipped with an AI and the remaining interferometers are clustered at the centre of the baseline.\footnote{These three configurations are illustrative and have not been informed by experimental constraints, which will ultimately dictate the layout of any multigradiometer experiment.}
 These three layouts are shown in Fig.~\ref{fig:configurations} for two, three and seven interferometers.
 
 In detail, in the unequal spacing (ends) configuration, the interferometers are clustered at either extremity of the baseline with the separation between nearest neighbours in each cluster determined by $L/2(\mathcal{N}-1)$. 
 The separation between nearest neighbours in each cluster is shorter than the shortest separation between interferometers in different clusters. When $\mathcal{N}$ is odd, one additional interferometer is in the bottom cluster relative to the cluster near the Earth's surface.
 In the equal spacing configuration, the separation between nearest neighbours is given by $L/(\mathcal{N}-1)$. 
 Finally, in the unequal spacing (centre) configuration, each extremity of the baseline is equipped with an AI and the remaining interferometers are clustered at the centre of the baseline. The separation between nearest neighbours in the cluster is given by $L/2(\mathcal{N}-1)$, which again implies that the separation between nearest neighbours in the central cluster is shorter than the shortest separation between interferometers in the cluster and at the ends of the baseline. In this case, the interferometers are positioned symmetrically around the baseline's midpoint.

\subsubsection{Searches below 1~Hz}

\begin{figure*}[t!]
    \centering
    \includegraphics[width=.49\columnwidth]{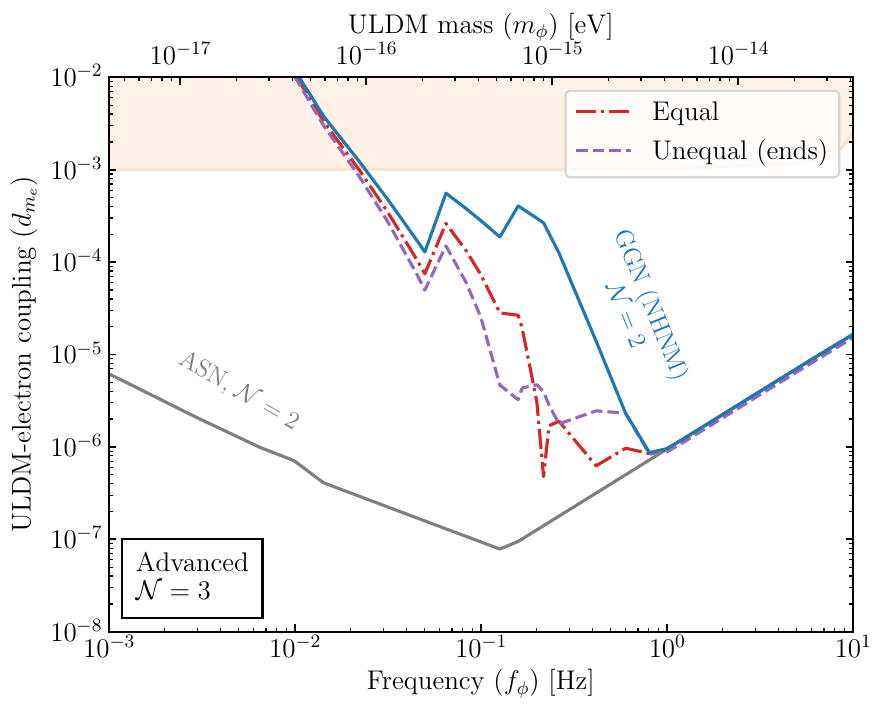}
\includegraphics[width=.49\columnwidth]{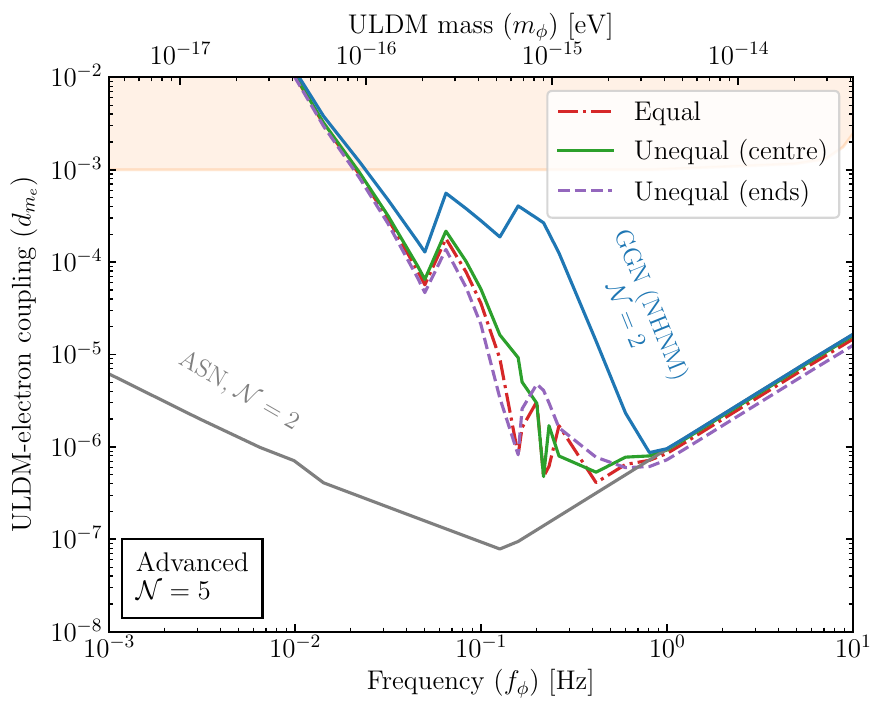}
    \caption{Projected 95\%~CL exclusion curves for an atom multigradiometer using the `advanced' parameters with $\mathcal{N}=3$ (left panel) and $\mathcal{N}=5$ (right panel) assuming GGN is modelled by the NHNM.
The grey and blue lines show the exclusion curve for a single atom gradiometer ($\mathcal{N}=2$) for ASN-only and in an ASN and GGN background, respectively.
The red dot-dashed, purple dashed and green solid lines show the atom multigradiometer exclusion curves for the equally-spaced, unequally-spaced (ends) and unequally-spaced (centre) configurations defined in Fig.~\ref{fig:configurations}, respectively. The multigradiometer exclusion curves regain parts of parameter space between $\sim 0.1$\,Hz and $\sim 1$\,Hz and have a similar sensitivity to the $\mathcal{N}=2$ configuration at higher frequencies. The orange shaded region is excluded by MICROSCOPE~\cite{PhysRevLett.129.121102}.}
    \label{fig:SensitivityAION1km_N35}
\end{figure*}

In Fig.~\ref{fig:SensitivityAION1km_N35} we show the 95\%~CL exclusion curves in the ULDM frequency (or mass) versus ULDM-electron coupling plane for different multigradiometer configurations, assuming the `advanced' design and under the assumption of the NHNM GGN scenario.
We begin by discussing the left panel, which compares the equal and unequal (ends) scenario for $\mathcal{N}=3$ with the single atom gradiometer ($\mathcal{N}=2$) configuration.\footnote{The equal spacing and unequal spacing (centre) configurations are identical for $\mathcal{N}=3$.} 
We observe an enhancement in the $\mathcal{N}=3$ exclusion curves at frequencies between $\sim 0.05$~Hz and $\sim 1$~Hz.
The enhancement has a mild dependence on the position of the interferometers along the baseline since, as explained in the previous section,
searches in the $\sim (0.05- 0.2)$~Hz frequency range favour the positioning of an interferometer pair towards the bottom of the baseline, while above $\sim 0.2$~Hz
equally-spaced configurations are preferred.

Below $\sim 0.05$~Hz, we see that the sensitivity is largely insensitive to the location of the interferometers. This is because at these low frequencies the GGN length scale is much larger than the baseline ($\lambda_{\rm{GGN}}\gg L$), so the GGN gradiometer phase is largely insensitive to the location of the interferometers.
Unlike in Fig.~\ref{fig:DiffDeltaz}, where the enhancement in the sub-Hz regime for a single atom gradiometer ($\mathcal{N}=2$) came at the expense of sensitivity loss above 1\,Hz, for $\mathcal{N}=3$ we find an exclusion curve above 1\,Hz that is comparable to the $\mathcal{N}=2$ configuration. We will postpone a more thorough discussion of the regime above 1\,Hz to the next sub-section.

The right panel of Fig.~\ref{fig:SensitivityAION1km_N35} compares the equal, unequal (ends) and unequal (centre) scenarios for $\mathcal{N}=5$ with the single atom gradiometer ($\mathcal{N}=2$) configuration. We see a similar pattern as in the left panel, with an enhancement at frequencies between $\sim0.05$~Hz and $\sim 1$~Hz. Again, we find a mild enhancement below $\sim 0.2$~Hz for the configuration that has the smallest $\Delta z$ towards the bottom of the baseline, namely, the unequal (ends) configuration.
Comparing the left and right panels, we observe a weak dependence (of at most a factor of a few) on the number of interferometers. This stems from the fact that $\mathcal{N} \gg 1$ AIs can simultaneously probe a larger array of length scales, which in turn means that the configurations will asymptotically approach the optimal gradiometer length and AI vertical position to maximise the reach across \textit{all} frequencies.
In this frequency range, the advantage of using $\mathcal{N}>3$ instead of $\mathcal{N}=3$ interferometers is minimal. 

Therefore, we have shown that since multigradiometer configurations can
 \textit{simultaneously} probe different length scales and vertical positions, they can achieve a frequency-dependent and weakly $\mathcal{N}$-dependent sensitivity enhancement over the single atom gradiometer ($\mathcal{N}=2$) configuration.

\subsubsection{Searches above 1~Hz}

At high frequencies, the background of all experiments considered in this work is dominated by atom shot noise. 
In contrast to the GGN and ULDM signals, atom shot noise is position-independent and is inversely proportional to the number of atoms $N_{\rm{atom}}$ employed in each interferometer sequence. 
As Fig.~\ref{fig:SensitivityAION1km_N35} demonstrates, the sensitivity above 1\,Hz is only weakly-dependent on the position of the interferometers along the baseline.
Therefore, in the regime where GGN can be neglected, we can estimate the sensitivity scaling with the number of interferometers by making the unphysical assumption that all interferometers are placed at the extreme ends of the baseline.
In this case, we expect a $\sqrt{\mathcal{N}}$ sensitivity scaling since increasing $\mathcal{N}$ is equivalent to increasing $N_{\rm{atom}}$ in each cloud, and we know that the sensitivity scales like $\sqrt{N_{\rm{atom}}}$ in the absence of GGN (cf.\ section~\ref{subsec:TSAsimov}).

Considering a more realistic multigradiometer configuration with equal spacing between the atom interferometers, we find analytically that the test statistic is related to the test statistic for a single gradiometer with maximum gradiometer length and identical parameters by
\begin{equation}
    q^{\mathrm{multi}}_\mathcal{N} = \frac{1}{9}\left (\frac{\mathcal{N}}{2}\right )^2\left (\frac{\mathcal{N}+1}{\mathcal{N}-1}\right )^2 q_{\mathcal{N}=2} \,.
    \label{eq: TS equal spacing}
\end{equation}
Given that the test statistic is proportional to the fourth-power of the ULDM coupling (cf.\ section~\ref{subsec:TSAsimov}), we find explicitly the $\sqrt{\mathcal{N}}$ scaling in the limit of large $\mathcal{N}$.
Setting $\mathcal{N}=3$ in Eq.~\eqref{eq: TS equal spacing}  shows that there is no sensitivity enhancement when an additional atom interferometer is positioned at the mid-point along the baseline.
This behaviour is the reason why the unequal spacing (centre) configurations, which feature an AI at either end of the baseline and equally spaced interferometers clustered around the baseline's mid-point, 
display a smaller sensitivity increase than equally-spaced interferometers. Conversely, the unequal spacing (ends) configuration provides the best sensitivity at high frequencies since the interferometers are furthest from the mid-point.

\subsubsection{Searches in the large $c_H$ regime}

In the projections shown until this point, we have utilised a set of geological parameters in which the GGN length scale $\lambda_{\rm{GGN}}$, defined in Eq.~\eqref{eq:GGNscalelength}, is similar in magnitude to the baseline at frequencies between $\sim 10^{-1}$~Hz and $\sim 1$~Hz.
This was achieved by considering a km-baseline experiment and a ground material in which the Rayleigh wave's horizontal speed was $c_H \sim250~\mathrm{m\,s^{-1}}$.
We now consider the scenario where $\lambda_{\rm{GGN}}\gtrsim L$. This would occur, for example, if the ground material consisted of limestone, for which $c_H \simeq3232~\mathrm{m\,s^{-1}}$ and $\lambda_{\rm{GGN}}\simeq\SI{1.3}{km}~\left(\frac{3232~\mathrm{m\,s^{-1}}}{ c_H}\right)^{-1}\,\left(\frac{2.5~\mathrm{Hz}}{\omega_a}\right)$.

\begin{figure*}[t]
    \centering
    \includegraphics[width=.49\columnwidth]{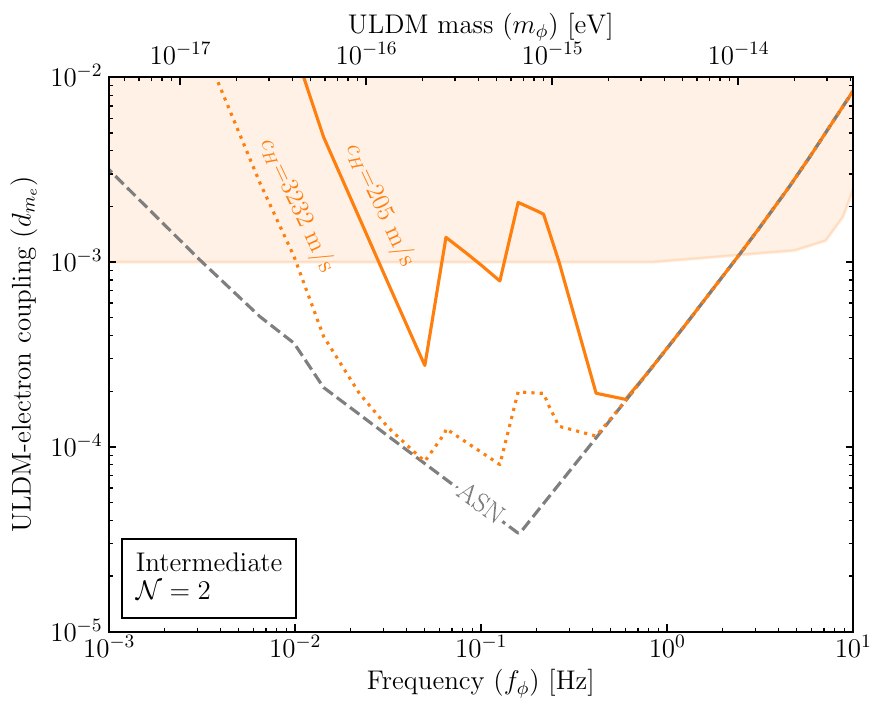} 
    \includegraphics[width=.49\columnwidth]{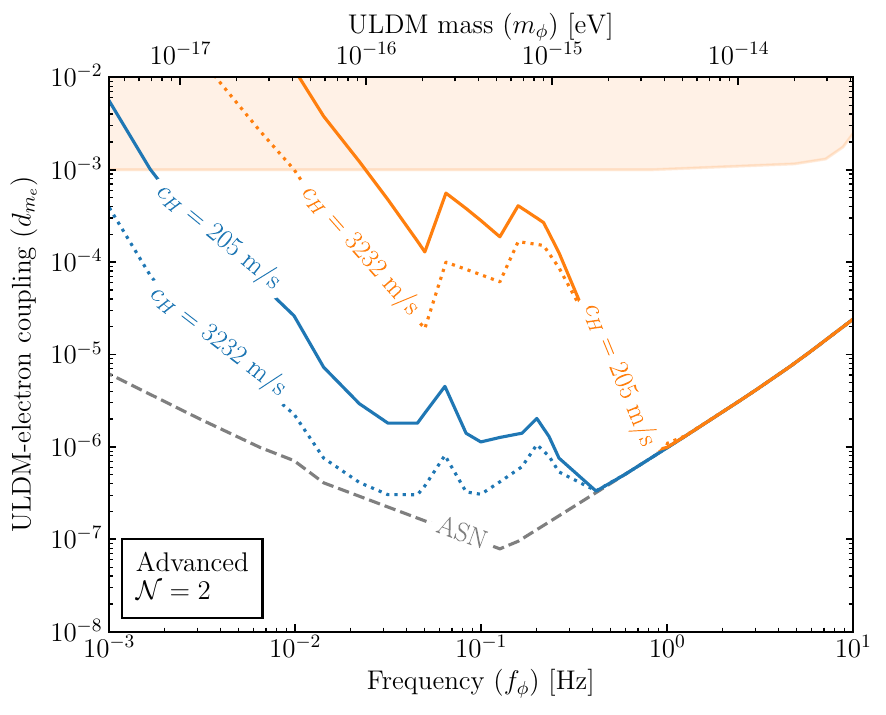}\\
    \includegraphics[width=.49\columnwidth]{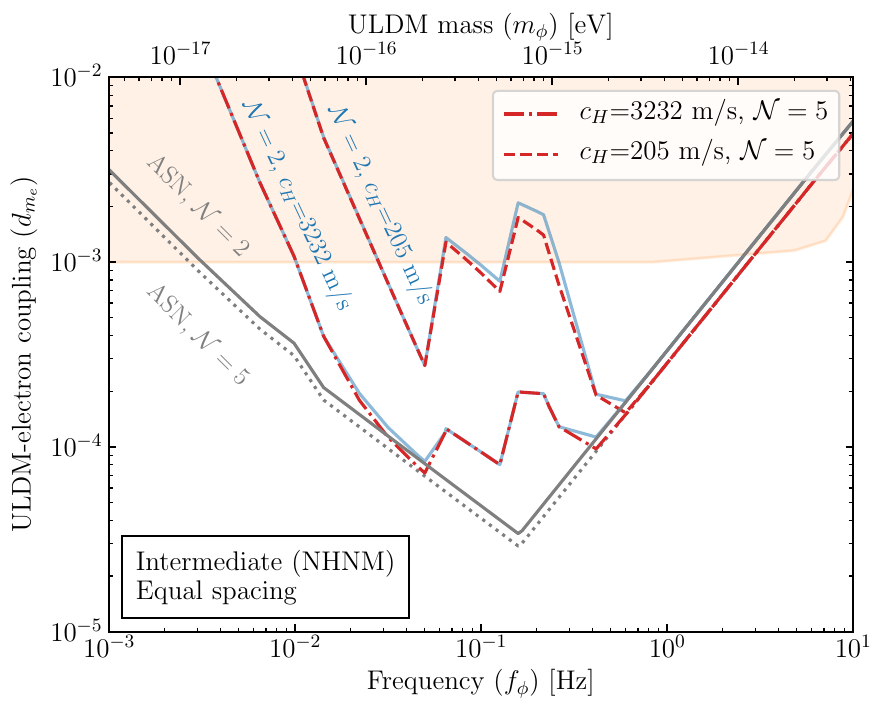}
    \includegraphics[width=.49\columnwidth]{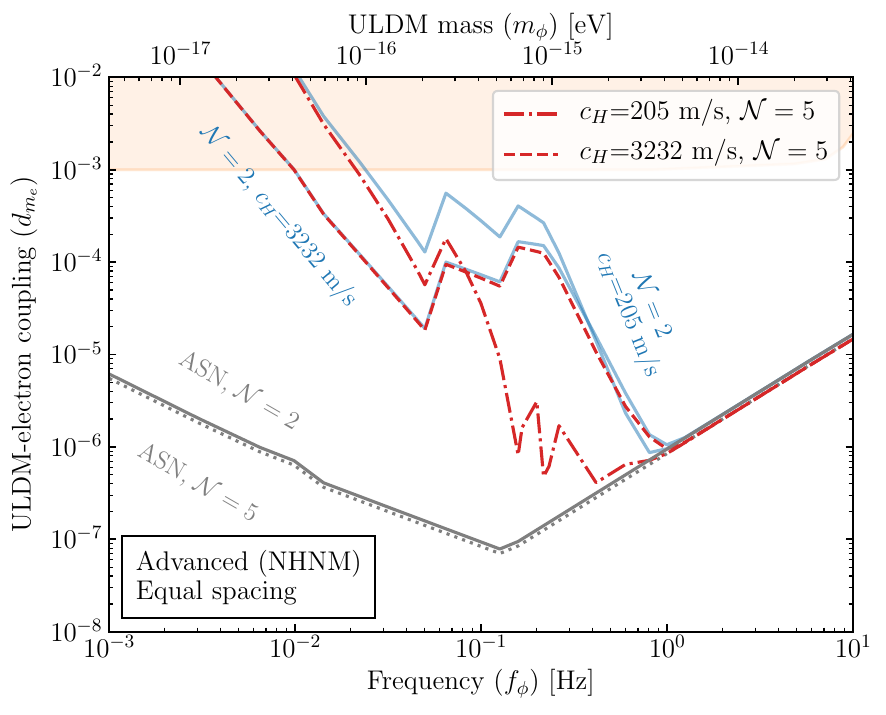}
    \caption{ 
    Impact of changing the Rayleigh wave's horizontal ground speed $(c_H)$ on the projected 95\% CL exclusion curves of atom gradiometers. 
    Upper panels: exclusion curves for a single atom gradiometer ($\mathcal{N}=2$) for the `intermediate' (left) and `advanced' (right) designs. Solid and dotted lines show $c_H = 205\,\mathrm{m\,s}^{-1}$ and $c_H = 3232\,\mathrm{m\,s}^{-1}$, respectively. The NHNM and NLNM (upper-right only) scenarios are shown in orange and blue, respectively.
    Lower panels: comparison of the exclusion curves for $\mathcal{N}=2$ (blue) and $\mathcal{N}=5$ (red) assuming that the AIs are equally spaced under the NHNM scenario. When $c_H = 3232$\,m/s, the $\mathcal{N}=5$ configuration provides no increase in sensitivity relative to $\mathcal{N}=2$. 
    In all panels the orange shaded region is excluded by MICROSCOPE~\cite{PhysRevLett.129.121102}. The solid grey lines show the exclusion curves assuming $\mathcal{N}= 2$ and an atom shot noise-only background, while the dotted grey lines show the exclusion curve assuming a $\mathcal{N}=5$ ASN-limited and equally-spaced multigradiometer experiment.} 
    \label{fig:cH comparison}
\end{figure*}

The upper panels of Fig.~\ref{fig:cH comparison} compare the sensitivity reach between the two values of the horizontal speed: $205~\mathrm{m\,s^{-1}}$, which we show with solid curves; and $3232~\mathrm{m\,s^{-1}}$, which we show with dotted curves. In the upper-left panel, we focus on a 100\,m baseline operating with $\mathcal{N}=2$ and the `intermediate' parameters from Table~\ref{table:ExperimentalParameters}. Here, we only display the projected reach in the NHNM scenario, since the NLNM curve is degenerate with the atom shot noise-limited curve. In the upper-right panel, we consider the scenario for a 1\,km baseline operating with $\mathcal{N}=2$ and the `advanced' parameters from Table~\ref{table:ExperimentalParameters}. Differently from the left panel, we display the sensitivity curves in both the NHNM and NLNM scenarios, which we plot in orange and blue respectively. Under both the NHNM and NLNM scenarios, the $c_H =3232~\mathrm{m\,s^{-1}}$ sensitivity curves are significantly lower than the $c_H =205~\mathrm{m\,s^{-1}}$ curves over all frequencies in the sub-Hz regime for both baselines, 
which suggests that geological materials with large $c_H$ values can naturally act as GGN filters owing to the smaller value of $\lambda_{\mathrm{GGN}}$.\footnote{There is  an 11\% difference in the ground density between the low and high $c_H$ scenarios but since the GGN phase scales linearly with the ground density (Eq.~\eqref{eq:app:GGN} and the discussion that follows), this cannot explain the  large change observed in Fig.~\ref{fig:cH comparison}.} This follows from the fact that in the regime where $\lambda_\mathrm{GGN} > L$, the GGN gradiometer phase shift is inversely proportional to $\lambda_\mathrm{GGN}$, and so also inversely proportional to $c_H$; hence, larger values of $c_H$ imply a smaller GGN gradiometer phase shift and, in the limit where the background is dominated by GGN, better sensitivity to linearly-coupled scalar ULDM.

The lower panels in Fig.~\ref{fig:cH comparison} compare the $\mathcal{N}=2$ configuration shown in blue with the $\mathcal{N}=5$ equal spacing multigradiometer configuration, shown in red dot-dashed and red dashed for $c_H=205~\mathrm{m\,s^{-1}}$ and $c_H =3232~\mathrm{m\,s^{-1}}$ respectively. In both lower panels we focus on the NHNM scenario.
In the lower-left panel, where we study the `intermediate' scenario with a 100\,m baseline, we observe that when GGN dominates the background (in the sub-Hz range), only a small sensitivity enhancement is observed for $\mathcal{N}=5$ compared to $\mathcal{N}=2$ when $c_H =205~\mathrm{m\,s^{-1}}$, and there is essentially no improvement when $c_H =3232~\mathrm{m\,s^{-1}}$. This follows because the GGN length scale of the Rayleigh wave's profile underground is already $\mathcal{O}(100)$\,m when $c_H =205~\mathrm{m\,s^{-1}}$ and significantly exceeds the length of the baseline when $c_H =3232~\mathrm{m\,s^{-1}}$, so the GGN gradiometer phase is largely insensitive to the location of the interferometers. 
In the km-baseline `advanced' scenario, shown in the lower-right panel, an $\mathcal{N}=5$ configuration significantly improves the sensitivity in the $\sim 0.1$\,Hz to $\sim 1$\,Hz  frequency range relative to $\mathcal{N}=2$ when $c_H =205~\mathrm{m\,s^{-1}}$. However, we again see  no such improvement for $c_H =3232~\mathrm{m\,s^{-1}}$, since at these frequencies $\lambda_\mathrm{GGN}\gg L$. This highlights the fact that in the $\lambda_\mathrm{GGN}\gg L$ regime, the multigradiometer configuration is a less useful tool to mitigate~GGN.

\section{Discussion and summary}\label{sec:Discussion}

Single-photon atom gradiometry is a powerful experimental technique that can be employed to search for the oscillation of atomic transition energies induced by ULDM.
By developing a robust statistical analysis based on a likelihood-based frequentist framework, we have provided a careful treatment that characterises the impact of GGN on ULDM searches with vertical atom gradiometers.
We have modelled the GGN as arising from an incoherent superposition of fundamental Rayleigh-wave modes, and used Peterson's data-driven models to capture the expected behaviour of low and high seismic-noise scenarios.
We find that GGN has the potential to significantly impact the sensitivity reach of longer-baseline experiments, $L\gtrsim\mathcal{O}(100\,~\mathrm{m})$, in the sub-Hz regime. 
However, we have shown that vertical atom multigradiometer experiments, consisting of  three or more atom interferometers in the same baseline, can recover significant parts of parameter space.
Our results demonstrate how robust sensitivity projections for vertical atom gradiometers, such as future versions of AION and MAGIS-100, can be obtained for frequencies down to $10^{-3}$~Hz, corresponding to an ULDM mass of $\sim 10^{-17}$~eV.

There is scope to go beyond the analysis presented in this work.
Firstly, the geological parameters that we used here have been chosen to be a generic representation, so they may not accurately describe the geology of the site where AION-100, MAGIS-100 or other future vertical atom gradiometers will be located. 
Secondly, the homogeneous half space model for the ground surrounding the experiment may not accurately describe more realistic geological settings.
Indeed, the model does not describe the existence of geological anisotropies or the presence of different geological strata.
For example, it may be the case that isotropy is broken by a body of water located close to the shaft, as would occur at the Boulby Underground Laboratory or at CERN owing to their vicinity to the North Sea and Lake Geneva, respectively.
Additionally, different geological strata would give rise to a much richer spectrum of Rayleigh modes. 
This is because
the additional strata would create more interfaces for the reflection and propagation of P- and S-waves,
and this would generate Rayleigh wave overtones beyond the fundamental mode.
Thirdly, in addition to the geological model, our calculations are dependent on the vertical displacement spectrum in the sub-Hz regime. 
A specific site may not follow either of Peterson's NLNM or NHNM scenarios, which attempt to bracket the range of possibilities. 
Therefore, a site-specific spectrum acquired as part of a dedicated site characterisation campaign would be preferred.
Finally, our model of GGN could be extended to include the contributions beyond seismic Rayleigh waves, such as atmospheric pressure-induced perturbations and seismic body waves that propagate through the Earth.
In summary, to further improve the sensitivity predictions in the sub-Hz regime for future versions of AION or MAGIS-100, the homogeneous half space model should be extended 
to account for the possibilities discussed above, and a detailed site-specific seismic and ground measurement campaign should be undertaken.
We emphasise that a key advantage of the likelihood-based analysis presented in this work is that the updated geological models and the site-specific information can be straightforwardly incorporated into the analysis procedure.

While our focus has been on scalar ULDM searches, 
we highlight that the likelihood-based frequentist framework developed here can be extended to stochastic gravitational wave searches with an atom multigradiometer experiment.
 A study exploring these signals in the mid-frequency band using this framework will be the focus of future work.

\begin{acknowledgments}
We are grateful to members of the AION Collaboration for many fruitful discussions and to John Ellis for comments on the manuscript. L.B. acknowledges support from the Science and Technology Council (STFC) Grant No. ST/T506199/1. C.M. acknowledges support from the Science and Technology Facilities Council (STFC) Grants No.~ST/T00679X/1 and No.~ST/N004663/1. J.M. acknowledges support from the University of Cambridge Isaac Newton Trust. We are grateful to the STFC for funding contributions towards the AION-10 and MAGIS-100 experiments. For the purpose of open access, the authors have applied a Creative Commons Attribution (CC BY) licence to any Author Accepted Manuscript version arising from this submission. No experimental datasets were generated by this research.
\end{acknowledgments}

\appendix

\section{ULDM signal phase \label{app:ULDMphase}}

A linearly-coupled scalar ULDM field $\phi(t)$, defined in Eq.~\eqref{eq:full-field}, which has interactions governed by the  Lagrangian 
\begin{equation}
\mathcal{L}_{\phi} \supset \phi(t) \sqrt{4 \pi G_{N}} \left[\frac{d_{e}}{4 e^{2}} F_{\mu \nu} F^{\mu \nu} -d_{m_{e}} m_{e} \overline{\psi}_e\psi_e \right],
\end{equation}
will induce oscillations in an atom's transition frequency (see e.g.,~\cite{Arvanitaki:2014faa, Stadnik:2014tta, Stadnik:2015kia}):
\begin{equation}
\omega_A(t)=\omega_A +  \overline{\Delta \omega_A}\, \sum_a \alpha_a \sqrt{F_\mathrm{DM}(v_a)}  \cos(\omega_a t + \theta_a)\;.
\end{equation}
Here, $d_e$ and $d_{m_e}$ are dimensionless and parameterise the coupling strength relative to the Planck mass $1/\sqrt{4\pi G_N}$, where  $G_N$ is Newton's gravitational constant, 
$\alpha_a$ is a Rayleigh distributed variable and $\theta_a$ is a random phase. The sum is carried over DM velocity classes of width $\Delta v$ that are weighted by the DM speed distribution.
For the $5\mathrm{s}^2 \,^1\mathrm{S}_0\leftrightarrow 5\mathrm{s}5\mathrm{p} \,^3\mathrm{P}_1$ clock transition 
in $^{87}\mathrm{Sr}$, which we assume throughout this work, $\omega_A=2.697\times10^{15}~\mathrm{rad}/\mathrm{s}$, while 
\begin{equation}
    \frac{\overline{\Delta \omega_A}}{\omega_A} =  d_{\phi} \sqrt{4\pi G_N} \frac{ \sqrt{\rho_{\mathrm{DM}}}}{m_{\phi}} \, ,
    \label{eq: delta omega_A}
\end{equation}
where the ULDM-Standard Model coupling
strength is
\begin{equation}
    d_{\phi} = d_{m_e} + (2+\xi_A) d_e\; ,
\end{equation}
and $\xi_A$ is a calculable parameter that takes the value $\xi_A\approx0.06$ for $^{87}\mathrm{Sr}$~\cite{Angstmann:2004zz}.

Following the procedure described in Ref.~\cite{Badurina:2021lwr},
the expression for the gradiometer phase in natural units measured at the end of the $m$\textsuperscript{th} launch between the $i$\textsuperscript{th} and $j$\textsuperscript{th} AIs, which are separated by the vertical distance $\Delta z^{(i,j)}$, can be expressed as
\begin{equation}\label{eq:app:ULDMphase}
   \Phi^{(i,j)}_{\mathrm{DM},m} = \frac{\Delta z^{(i,j)}}{L} \sum_{a} \alpha_a \sqrt{F_\mathrm{DM}(v_a)} \, A_a \cos \phi_{a,m}\;,
\end{equation}
where we have defined
\begin{gather}
\begin{split}\label{eq:app:ULDM amp}
    A_a = 8 \, \frac{\overline{\Delta \omega_A}}{\omega_a} & \sin\left[\frac{\omega_{a}n L}{2}\right] \sin \left[\frac{\omega_{a}(T-(n-1)L)}{2} \right]  \sin\left[\frac{\omega_{a} T}{2}\right]\;, 
    \end{split} \\
        \phi_{a,m} = \omega_{a}  \left ( \frac{2T+L}{2} + m\Delta t \right )+ \theta_a\;,
\end{gather}
and, as previously stated, $L$ is the baseline, $n$ is the number of LMT kicks, while $T$ is the interrogation time. For definiteness, the random variables that enter Eq.~\eqref{eq:app:ULDMphase} are $\alpha_a$ and~$\phi_{a,m}$.

\subsection{Statistical properties \label{app:ULDMphaseStatisticalProperties}}

The statistical properties of the ULDM signal follow straightforwardly from the statistical properties of the ULDM wave. The expectation value of the ULDM phase shift~is
\begin{equation}
\begin{aligned}
\Big \langle \Phi^{(i,j)}_{\mathrm{DM},m} \Big \rangle = \frac{\Delta z^{(i,j)}}{L} \sum_{a} \Big \langle \alpha_a \sqrt{F_\mathrm{DM}(v_a)} \, A_a \cos \phi_{a,m} \Big \rangle  \, .
\end{aligned}
\end{equation}
Since the random phase $\theta_a$ and the Rayleigh distributed random variable $\alpha_a$ are independent, the expectation value will act on the cosine term and separately on the amplitude, such that
\begin{equation}
\begin{aligned}
\Big \langle \Phi^{(i,j)}_{\mathrm{DM},m} \Big \rangle = \frac{\Delta z^{(i,j)}}{L} \sum_{a}  \langle \alpha_a \rangle \sqrt{F_\mathrm{DM}(v_a)} \, A_a \langle \cos \phi_{a,m} \rangle \, .
\end{aligned}
\end{equation}
According to the definition of the Rayleigh distribution, $\langle \alpha_a \rangle > 0$; however, since $\phi_{a, m}$ depends linearly on the random phase $\theta_a$, $\langle \cos \phi_{a,m} \rangle$ vanishes, such that
\begin{equation}
\begin{aligned}
\Big \langle \Phi^{(i,j)}_{\mathrm{DM},m} \Big \rangle = 0 \, .
\end{aligned}
\end{equation}

The covariance of the gradiometer also follows from the statistical properties of the ULDM wave. Neglecting for brevity the factors depending on the gradiometer length, using the vanishing expectation value of the ULDM phase and recalling the fact that $\alpha_a$ and $\theta_a$ are independent random variables, the expected covariance is proportional to
\begin{equation}
\begin{split}
\Big \langle \Phi^{(i,j)}_{\mathrm{DM}, m} \Phi^{(i',j')}_{\mathrm{DM}, m'}\Big \rangle  \propto & \sum_{a, a'} \langle \alpha_a \alpha_{a'} \rangle \sqrt{F_\mathrm{DM}(v_a)F_\mathrm{DM}(v_{a'})} \\ &\times \, A_a A_{a'} \langle \cos \phi_{a,m} \cos \phi_{a',m'} \rangle \, .
\end{split}
\end{equation}
Since $\phi_{a,m}$ depends on a random phase $\theta_a$, the sum over the velocity classes (labelled by $a$ and $a'$) is non-zero only when the random phases are the same, i.e. $a'=a$. Hence, we may write
\begin{equation}
\begin{split}
\Big \langle \Phi^{(i,j)}_{\mathrm{DM}, m} \Phi^{(i',j')}_{\mathrm{DM}, m'}\Big \rangle 
& \propto  \sum_{a} \langle \alpha_a^2 \rangle F_\mathrm{DM}(v_a)  \, A_a^2 \langle \cos \phi_{a,m} \cos \phi_{a,m'} \rangle \, .
\end{split}
\label{eq:app:ULDMCovStep}
\end{equation}
To further simplify the expression above, we observe that $\left \langle \cos \phi_{a,m} \cos \phi_{a,m'}\right \rangle$ can be re-expressed in terms of double angle formulas as
\begin{equation}
    \begin{split}
    \left \langle \cos \phi_{a,m} \cos \phi_{a,m'}\right \rangle &= \frac{1}{2}\left [ \left\langle \cos \left ( \phi_{a,m} - \phi_{a,m'} \right ) \right.\right.  + \left. \left. \cos \left ( \phi_{a,m} + \phi_{a,m'} \right )\right \rangle \right] \\
    & = \frac{1}{2}  \cos \left ( \omega_{a} \Delta t (m-m') \right ) + \frac{1}{2} \left \langle  \cos \left ( \phi_{a,m} + \phi_{a,m'} \right ) \right \rangle \\
    & = \frac{1}{2} \cos \left ( \omega_{a} \Delta t (m-m') \right )\,,
    \end{split}
\label{eq:app:CosProperty}
\end{equation}
where the second line follows from the definition of $\phi_{a, m}$ and the last line from the vanishing expectation value of the cosine of a random phase. Using this result and $\langle \alpha_a^2 \rangle$ = 2, and also restoring the dependence on the gradiometer length, we find that the covariance takes the form
\begin{equation}\label{eq:app:ULDMCov}
\begin{split}
\Big \langle \Phi^{(i,j)}_{\mathrm{DM}, m} \Phi^{(i',j')}_{\mathrm{DM}, m'}\Big \rangle 
& =\frac{\Delta r^{(i,j)}}{L} \frac{\Delta r^{(i',j')}}{L}\\
&\times \sum_{a} F_\mathrm{DM}(v_a) \, A_a^2 \cos (\omega_a \Delta t(m-m')) \, .
\end{split}
\end{equation}

\section{GGN from the fundamental Rayleigh mode \label{app:RayleighGGN}}

Our GGN model assumes that the dominant contribution arises from the fundamental Rayleigh mode. The fundamental Rayleigh mode consists of S-waves with vertical displacements that are coupled to P-waves. In this appendix, we begin by deriving an expression for the angle-averaged gravitational potential caused by the fundamental Rayleigh mode. This potential is used as an input to calculate the phase shift induced in an atom interferometer. We then consider the statistical properties of the phase shift. Our notation largely follows Refs.~\cite{hughes_seismic_1998,Landau:1986aog}, and we also make extensive use of the discussion in Ref.~\cite{Harms:2015zma}.

We begin with some definitions. The propagation speed of P-waves,~$c_P$, is determined by the material’s density $\rho_0$, and bulk and shear moduli $K$ and~$\mu$, respectively,~as
\begin{equation}
    c_{P}=\sqrt{\frac{K+4 \mu / 3}{\rho_0}} \, .
\end{equation}
The propagation speed of S-waves, $c_S$, is related to that of P-waves by
\begin{equation}
    c_{S}=\sqrt{\frac{1-2 \nu}{2-2 \nu}}\, c_{P}\;,
\end{equation}
where $\nu$ is the material’s Poisson ratio, which depends on the ground's bulk and shear moduli as
\begin{equation}
    \nu=\frac{3 K-2 \mu}{2(3 K+\mu)} \, .
\end{equation}

We model the geological surroundings of the experimental site as a homogeneous half space with constant ground density $\rho_0$ and Poisson ratio $\nu$ across all depths. This is an idealised scenario which implies that the Rayleigh waves are isotropic and that only the fundamental Raleigh mode is present: Rayleigh overtones require geologically stratified structures, which are not present in our model. The fundamental Raleigh mode has frequency~$f_a$, angular frequency~$\omega_a = 2\pi f_a$, horizontal wave number~$k_a$, horizontal propagation speed~$c_H = \omega_a/k_a$, and horizontal propagation direction~$\boldsymbol{\hat{k}}$. The horizontal propagation speed~$c_H$ is related to $c_S$ by $c_H = c_S \sqrt{\zeta}$, where $\zeta$ is the real root of the cubic equation
\begin{equation}
\label{eq:cubic}
\zeta^{3}-8 \zeta^{2}+8\left(\frac{2-\nu}{1-\nu}\right) \zeta-\frac{8}{(1-\nu)}=0\;.
\end{equation}
It is convenient to introduce the dimensionless ratios $s$ and $q$,
\begin{align}
    s&=\sqrt{1-\left(c_{H} / c_{S}\right)^{2}}\;, \\
     q&=\sqrt{1-\left(c_{H} / c_{P}\right)^{2}} \;,
\end{align}
which will appear as the vertical e-folding rate to the horizontal wave number for S-waves and as the vertical e-folding rate to the horizontal wave number for P-waves, respectively. Equation~\eqref{eq:cubic} encodes the relation
\begin{equation}
1+s^2=2\sqrt{qs}\;,
\end{equation}
which we make extensive use of in the manipulations below.

Finally, the definition of our coordinate system is shown in Fig.~\ref{fig: GGN}.  It is convenient to express all vectors in cylindrical coordinates,
\begin{align}
    \boldsymbol{x} &= (\varrho \cos \theta,\varrho\sin \theta,0)\, ,\\
        \boldsymbol{r} &= (\varrho\cos \theta,\varrho\sin \theta,z)\;, \\
       \boldsymbol{k} &=  (k\cos \theta',k \sin \theta',0)\,.
\end{align}
In this system, $\boldsymbol{x}$ defines a point on the horizontal plane, and $z$ defines the depth into the ground. The product of the horizontal wave number and horizontal distance, $\boldsymbol{k} \cdot \boldsymbol{x}$, may be written as
\begin{equation}
    \boldsymbol{k} \cdot \boldsymbol{x} = k\varrho(\cos \theta' \cos \theta + \sin \theta' \sin \theta) = k\varrho \cos (\theta-\theta')\;.
\end{equation}

Having defined our terms, we now proceed to calculate the surface and underground density perturbations induced by the Rayleigh waves.
Adapting the expressions from Ref.~\cite{Harms:2015zma} to our coordinate system, the Rayleigh wave displacement vector is given by
\begin{equation}
\begin{split}
    \boldsymbol{\xi}_{a, R}(\varrho,\theta,z, t)&=\left(  \xi_{a, H}(z) \boldsymbol{\hat{k}}-  \xi_{a}(z) \boldsymbol{\hat{z}}\right) \\
    &\times \exp \left [{i(k_a\varrho \cos (\theta-\theta')-\omega_a t + \widetilde{\theta}_a)}\right ] \, ,
\end{split}
\end{equation}
where $\widetilde{\theta}_a$ is a random phase, $\boldsymbol{\hat{z}}$ is the unit vector pointing downward and
\begin{align}
\begin{split}
    \xi_{a}(z) &=  \xi_{a} \left(\frac{1+s^2}{1-s^2} \right) \Big[\exp {(-q \omega_a z/c_H)} - \frac{2}{1+s^2}\exp {(-s \omega_a z/c_H)}\Big] \, ,
    \end{split}
    \\
    \begin{split}
    \xi_{a, H}(z) &=   \frac{i \xi_{a}}{q} \left(\frac{1+s^2}{1-s^2} \right) \Big[\exp {(-q \omega_a z/c_H)}  -\frac{2 qs}{1+s^2}\exp {(-s \omega_a z/c_H)}\Big] \;,
    \end{split}
\end{align}
where $\xi_{a}$ is the vertical displacement at the surface, and the factor $i$ in $\xi_{a, H}(z)$ ensures that particle motion is retrograde at the surface. Thus, we see explicitly that the amplitude is characterised by an exponentially decaying profile while moving away from the surface.

The displacement field induced by the fundamental Rayleigh wave produces a fractional density perturbation at depth $z$~\cite{hughes_seismic_1998},
 \begin{equation}
 \begin{split}
     \frac{\delta \rho_a (z>0)}{\rho_0} &= \left[\xi_{a} \delta(z)+\mathcal{R}_a(z)\right] \exp \left [{i(k_a\varrho \cos (\theta-\theta')-\omega_a t + \widetilde{\theta}_a)}\right ]\, ,
     \end{split}
 \end{equation}
 where the first term is the surface contribution and arises from the displacement normal to the surface ($\delta(z)$ is the Dirac delta function), while the second term is the contribution from inside the medium and follows from the ground-mass continuity equation, such that
 \begin{align}
    \mathcal{R}_a(z) &=  -i \frac{\omega_a}{c_H} \xi_{a, H}(z) + \frac{\partial \xi_{a}(z)}{\partial z}\\
    &= \frac{\omega_a}{c_H}\xi_a \frac{(q^2-1)}{q} \left (\frac{1+s^2}{1-s^2} \right ) \exp {(-q\omega_a z/c_H)}\;.
 \end{align}
Following Ref.~\cite{Harms:2015zma}, the angle-averaged perturbing gravitational potential induced by a fundamental Rayleigh mode of frequency $\omega_a$, and experienced by a test mass underground at time $t$ and position $\boldsymbol{r}_0 = (0,0,z_0)$, is

 \begin{equation}
  \begin{split}
     V_a\left(z_0, t\right) = & - 2\pi G \rho_0\, \cos (\omega_a t + \widetilde{\theta}_a)\, \\ & \quad \times \int \varrho \mathrm{d} \varrho \,\mathrm{d} z \, \frac{1}{\sqrt{\varrho^2 +(z-z_0)^2}} \left[\xi_{a} \delta(z)+\mathcal{R}_a(z)\right] J_0(\omega_a\varrho/c_H) \, ,
\end{split}
 \end{equation}
 where $J_0(\omega_a\varrho/c_H)$ is the Bessel function of first kind.
 Performing the integrals, and summing over all frequencies, we arrive at the expression 
  \begin{equation}
    \begin{split}
     V_a\left(z_0, t\right) = & -2\pi G \rho_0\,\sum_a \xi_a \,\frac{c_H}{q\, \omega_a}\, \left(\frac{1+s^2}{1-s^2} \right)\\ & \times \left[ \left(1+\sqrt{q/s}\right) \exp {(-\omega_a z_0/c_H)} -2 \exp {(-q \omega_a z_0/c_H)}  \right] \cos (\omega_a t + \widetilde{\theta}_a) \, .
 \end{split}
 \end{equation}

\subsection{Rayleigh wave signal phase}\label{app:Rayleighphase}

The leading order phase shift induced by the perturbing potential on the atom interferometer can be calculated using the perturbative approach~\cite{Storey:1994oka}, in which the GGN potential is treated as a perturbation to Minkowski space. In flat space time, the atom trajectories follow straight lines, and the interferometer can be geometrically thought of as a kite in spacetime. In this geometry, the separation, laser, and propagation phases are all zero. Using the standard approach for perturbative atom interferometer calculations~\cite{Storey:1994oka}, the leading order phase shift from the fundamental Rayleigh wave field with angular frequency $\omega_a$ leaves a non-zero contribution to the propagation phase. In all experimental configurations considered here, we take the launch velocity $v_\mathrm{in}$ and velocity kick from the $n$ LMT pulses $v_\mathrm{kick}$ to be much smaller than the horizontal speed of the Rayleigh wave, such that $v_\mathrm{kick}/c_H$, $v_\mathrm{in}/c_H \ll 1$. Additionally, for all frequencies considered in this paper, the Rayleigh wave's wavelength is significantly larger than the length scales probed by the atoms during interferometry, i.e.\ $\omega_a T v_\mathrm{kick}/c_H \ll 1$, such that the leading order GGN-induced phase shift measured by the $i$\textsuperscript{th} AI at the end of the $m$\textsuperscript{th} launch takes the form 
\begin{equation}\label{eq:app:GGN}
    \Phi^{(i)}_{\mathrm{GGN},m} = \sum_{a}  \xi_a \Bigg (\widetilde{A}_a \exp \left (-q\frac{\omega z_i}{c_H} \right) + \widetilde{B}_a \exp \left ( -\frac{\omega z_i}{c_H} \right) \Bigg ) \cos{\widetilde{\phi}_{a,m}}\;.
\end{equation}
In this expression we have defined
\begin{gather}
\widetilde{A}_a = n k_A \, \frac{16\pi G \rho_0}{ \omega_a^2} \left(\frac{1+s^2}{1-s^2} \right) \sin^2 \left(\frac{\omega_a T}{2}\right) \, , \\
\widetilde{B}_a =- n k_A\, \frac{8\pi G \rho_0 }{q\, \omega_a^2}  \left(\frac{1+s^2}{1-s^2} \right) \,   \left(1+\sqrt{q/s}\right) \sin^2 \left(\frac{\omega_a T}{2}\right) \, , \\
\widetilde{\phi}_{a,m} =  \omega_a \left( T  +m \, \Delta t \right)+\widetilde{\theta}_a\, ,
\end{gather}
and assumed that the momentum kick from the LMT sequence is $n k_A$. For definiteness, the independent random variables that enter Eq.~\eqref{eq:app:GGN} are $\xi_a$ and $\widetilde{\theta}_a$.

With our assumption that the ground composition is homogeneous down the baseline, it follows that the GGN gradiometer phase between the $i$\textsuperscript{th} and $j$\textsuperscript{th} interferometers~is

\begin{equation}
\begin{split}
    \Phi^{(i,j)}_{\mathrm{GGN},m} =& \sum_{a}  \xi_a  \Bigg \{\widetilde{A}_a \left [ \exp \left (-q\frac{\omega z_i}{c_H} \right)-\exp \left (-q\frac{\omega z_j}{c_H} \right) \right] \\ & \quad + \widetilde{B}_a \left [ \exp \left ( -\frac{\omega z_i}{c_H} \right) - \exp \left ( -\frac{\omega z_j}{c_H} \right ) \right] \Bigg \} \cos{\widetilde{\phi}_{a,m}}\;.
\end{split}    
    \label{eq:app:GGNgradphase}
\end{equation}

In Figs.~\ref{fig:SensitivityAION100_N2}, \ref{fig:SensitivityAION1km_N2}, \ref{fig:DiffDeltaz} and \ref{fig:SensitivityAION1km_N35} we use the following ground parameters: $\rho_0 = 1800$~kg/m\textsuperscript{3}, $\nu = 0.33$, $c_P = 440$~m/s and $c_S=220$~m/s. From these, we find $s = 0.36$, $q = 0.88$ and $c_H = 205$~m/s.
In Fig.~\ref{fig:cH comparison} we also show results with $\rho_0 = 2000$~kg/m\textsuperscript{3}, $\nu = 0.34$, $c_P = 6964$~m/s and $c_S=3464$~m/s, which lead to $s = 0.36$, $q = 0.89$ and $c_H = 3232$~m/s.
We stress that these parameters are representative estimates and that site-specific measurements would be necessary to deduce these parameters.

\subsection{Statistical properties \label{app:RayleighphaseStatisticalProperties}}

Since the GGN phase is sourced by a stochastic superposition of waves, it follows that the statistical properties of the ULDM signal also characterise the GGN background contribution. Indeed, since the vertical displacement $\xi_a$ is Rayleigh distributed at each angular frequency $\omega_a$ and, importantly, is independent of the random phase $\theta_a$, the expectation value of the GGN gradiometer phase takes the form

\begin{equation}
\begin{split}
    \Big \langle \Phi^{(i,j)}_{\mathrm{GGN},m} \Big \rangle &= \sum_{a} \langle \xi_a \rangle \Bigg \{\widetilde{A}_a \left [ \exp \left (-q\frac{\omega z_i}{c_H} \right)-\exp \left (-q\frac{\omega z_j}{c_H} \right) \right] \\ &+ \widetilde{B}_a \left [ \exp \left ( -\frac{\omega z_i}{c_H} \right) - \exp \left ( -\frac{\omega z_j}{c_H} \right ) \right] \Bigg \} \langle \cos{\widetilde{\phi}_{a,m}} \rangle  \\& = 0 \, ,
\end{split}
\end{equation}
where the last equality follows from the fact that $\widetilde{\phi}_{a,m}$ depends linearly on a random phase~$\widetilde{\theta}_a$.

The calculation of the covariance also follows identically from the ULDM case. The only structural difference arises from the product of the exponential functions in Eq.~\eqref{eq:app:GGNgradphase}. Hence, using Eqs.~\eqref{eq:app:ULDMCovStep}-\eqref{eq:app:CosProperty}, it can easily be shown that the covariance is then given by
\begin{equation}
\begin{aligned}
\Big \langle \Phi^{(i,j)}_{\mathrm{GGN}, m} \Phi^{(i',j')}_{\mathrm{GGN}, m'}\Big \rangle 
& =\sum_{a}  \frac{\langle \xi_a^2 \rangle}{2} F_{\mathrm{GGN},a}^{(i,j)}F_{\mathrm{GGN},a}^{(i',j')}\cos (\omega_a \Delta t(m-m'))  \, ,
\end{aligned}
\label{eq:app:GGNCovariance}
\end{equation}
where 
\begin{gather}
        F_{\mathrm{GGN},a}^{(i,j)} = F_{\mathrm{GGN},a}^{(i)}-F_{\mathrm{GGN},a}^{(j)} \, , \\ F_{\mathrm{GGN},a}^{(i)} = \widetilde{A}_a \exp \left (-q\frac{\omega_a z_i}{c_H} \right) + \widetilde{B}_a \exp \left ( -\frac{\omega_a z_i}{c_H} \right) \, .
\end{gather}
It should be noted that, differently from $\alpha_a$, $\langle \xi_a^2 \rangle \neq 2$; hence, we obtain a term $\langle \xi_a^2 \rangle/2$ that is absent from Eq.~\eqref{eq:app:ULDMCov}. This is because $\alpha_a$ is a dimensionless random variable described by Eq.~\eqref{eq: Rayleigh distribution}, such that $\langle \alpha_a^2 \rangle  = 2$, while $\xi_a$ is not.

\section{Covariance matrices}\label{app:CovarianceMatrices}

 The covariance matrix of a multi-dimensional data vector $\mathbf{d}_k$ with zero expectation value is simply expressed as $\mathbf{\Sigma}_k  = \langle \mathbf{d}_k \mathbf{d}_k^\mathrm{\mathbf{T}} \rangle$. We remind the reader that the $\mathcal{N}(\mathcal{N}-1)$-dimensional data vector is defined as
\begin{equation}
    \mathbf{d}_k = \Big [ R_k^{(1,2)}, I_k^{(1,2)},\ldots , R_k^{(n, n-1)}, I_k^{(n,n-1)}\Big]^{\mathrm{\mathbf{T}}}
    \label{eq:app:DataVector}
\end{equation} 
in terms of the real and imaginary parts of the discrete Fourier transform (DFT), which we can explicitly write out in terms of noise and signal phase shifts measured by AG-($i,j$) as
\begin{align}
    R_{k}^{(i,j)}&=\frac{\Delta t}{\sqrt{T_\mathrm{int}}} \sum_{m=0}^{N-1}\left(\Phi^{(i,j)}_{\mathrm{Signal},m} + \Phi^{(i,j)}_{\mathrm{Noise},m}\right) c_{m, k} \\&= R_{k, \mathrm{Signal}}^{(i,j)}+R_{k, \mathrm{Noise}}^{(i,j)}\;, \label{eq:app:RealPart} \\
    I_{k}^{(i, j)}&=-\frac{\Delta t}{\sqrt{T_\mathrm{int}}} \sum_{m=0}^{N-1}\left(\Phi^{(i,j)}_{\mathrm{Signal},m} + \Phi^{(i,j)}_{\mathrm{Noise},m}\right) s_{m, k} \\&= I_{k, \mathrm{Signal}}^{(i,j)}+I_{k, \mathrm{Noise}}^{(i,j)}\;.
    \label{eq:app:ImaginaryPart}
\end{align}
For brevity, in Eqs.~\eqref{eq:app:RealPart}-\eqref{eq:app:ImaginaryPart} we use the shorthand notation for trigonometric functions, namely
\begin{equation}
    c_{m, k}=\cos \left(\frac{2 \pi k m}{N}\right) \; \text{ and }\;
    s_{m, k}=\sin \left(\frac{2 \pi k m}{N}\right) \, .
\end{equation}
Equipped with Eq.~\eqref{eq:app:DataVector} the covariance matrix of the $k$-frequency component collected by all gradiometers can then be organized into a symmetric $(\mathcal{N}(\mathcal{N}-1) \times \mathcal{N}(\mathcal{N}-1))$-dimensional matrix taking the form
\begin{equation}
\boldsymbol{\Sigma}_{k}=\left[\begin{array}{cccc}
\left\langle R^{(1,2)}_{k} R^{(1,2)}_{k}\right\rangle & \left\langle R^{(1,2)}_{k}I^{(1,2)}_{k}\right\rangle & \ldots & \left\langle R^{(1,2)}_{k} I^{(m-1,m)}_{k}\right\rangle \\
\left\langle I^{(1,2)}_{k} R^{(1,2)}_{k}\right\rangle & \left\langle I^{(1,2)}_{k}I^{(1,2)}_{k}\right\rangle  & \ldots & \left\langle I^{(1,2)}_{k} I^{(m-1,m)}_{k}\right\rangle \\
\vdots & & \ddots & \vdots \\
\left\langle I^{(m-1,m)}_{k} R^{(1,2)}_{k}\right\rangle & \left\langle I^{(m-1,m)}_{k}I^{(1,2)}_{k}\right\rangle & \ldots & \left\langle I^{(m-1,m)}_{k} I^{(m-1,m)}_{k}\right\rangle 
\end{array}\right] \, .
\label{eq:app:CovarianceMatrix}
\end{equation}

From Eq.~\eqref{eq:app:CovarianceMatrix}, it is clear that each entry in the covariance matrix consists of one of three unique products of imaginary and/or real parts, namely
\begin{equation}
    \left\langle R^{(i,j)}_{k} R^{(i',j')}_{k}\right\rangle, \, \left\langle R^{(i,j)}_{k} I^{(i',j')}_{k}\right\rangle, \, \left\langle I^{(i,j)}_{k} I^{(i',j')}_{k}\right\rangle \, .
\end{equation}
Additionally, from Eqs.\eqref{eq:app:RealPart}-\eqref{eq:app:ImaginaryPart}, each of these entries will contain terms pairing phases originating from identical or different sources. For example, using Eq.~\eqref{eq:app:RealPart} and without loss of generality, we find that
\begin{equation}
\begin{aligned}
    \left\langle R^{(i,j)}_{k} R^{(i',j')}_{k}\right\rangle &= \left\langle R^{(i,j)}_{k, \mathrm{Signal}} R^{(i',j')}_{k, \mathrm{Signal}}\right\rangle+\left\langle R^{(i,j)}_{k, \mathrm{Noise}} R^{(i',j')}_{k, \mathrm{Noise}}\right\rangle \\ & + \left\langle R^{(i,j)}_{k, \mathrm{Noise}} R^{(i',j')}_{k, \mathrm{Signal}}\right\rangle + \left\langle R^{(i,j)}_{k, \mathrm{Signal}} R^{(i',j')}_{k, \mathrm{Noise}}\right\rangle\,,
    \end{aligned}
\end{equation}
where the last two terms mix the signal and noise phases. Under our working assumption of uncorrelated noise and signal contributions, these terms vanish. Indeed, using the expectation value of the models considered in this work, we find
\begin{equation}
\begin{split}
    \left \langle R^{(i,j)}_{k, \mathrm{Noise}} R^{(i',j')}_{k, \mathrm{Signal}}\right\rangle &\propto \left \langle \Phi^{(i,j)}_{\mathrm{Noise},m} \Phi^{(i',j')}_{\mathrm{Signal},m'}\right \rangle \\&=  \left \langle \Phi^{(i,j)}_{\mathrm{Noise},m} \right \rangle \left \langle  \Phi^{(i',j')}_{\mathrm{Signal},m'}\right \rangle = 0\,.
    \end{split}
\end{equation}
The fact that the signal and each noise source are uncorrelated across the frequency spectrum allows us to decompose the covariance matrix $\boldsymbol{\mathrm{\Sigma}}_{k}$ into a signal $\boldsymbol{\mathrm{S}}_{k}$ and noise or background $\boldsymbol{\mathrm{B}}_{k}$ contribution. Assuming that each noise contribution is uncorrelated, we can also decompose the noise covariance matrix $\boldsymbol{\mathrm{B}}_{k}$ in terms of the GGN and atom shot noise contributions.
In the following subsections we will obtain analytical expressions for these covariance matrices.

\subsection{ULDM covariance matrix}\label{app:ULDMcovariancematrix}

The covariance matrix for the ULDM matrix can be calculated by recalling the statistical properties of the ULDM gradiometer phase. Focusing on the covariance between the real parts of the data's DFT, 
\begin{equation}\label{eq:app:RDMRDM}
\begin{split}
    \left\langle R^{(i,j)}_{k} R^{(i',j')}_{k}\right\rangle &= \frac{(\Delta t)^2}{T_\mathrm{int}} \sum_{m,m'=0}^{N-1}\left \langle \Phi^{(i,j)}_{\mathrm{DM},m}\Phi^{(i',j')}_{\mathrm{DM},m'}\right \rangle c_{m, k}\, c_{m', k} \\
    &= \frac{(\Delta t)^2}{T_\mathrm{int}} \frac{\Delta z^{(i,j)}}{L} \frac{\Delta z^{(i',j')}}{L} \sum_{m,m'=0}^{N-1} \sum_{a} F_\mathrm{DM}(v_a) A_a^2 \cos \left ( \omega_{a}\Delta t (m-m') \right ) c_{m, k} c_{m', k} \, ,
\end{split}
\end{equation}
where the second equality follows from the covariance of the ULDM gradiometer phase, shown in Eq.~\eqref{eq:app:ULDMCov}.
To make further progress, we recall that the ULDM speed indices $a$ can be mapped to those labelling the discrete frequencies probed by the experiment. Hence, we may rewrite the sum over $m$ and $m'$ as
\begin{equation}
    \begin{split}
        \sum_{m,m'=0}^{N-1} \cos \left ( \omega_{a}\Delta t (m-m') \right ) c_{m, k} c_{m', k} &=  \sum_{m,m'=0}^{N-1} (c_{m,k'}c_{m',k'}+s_{m,k'}s_{m',k'})c_{m,k}c_{m',k}\,,
    \end{split}
    \label{eq:app:CosineExpansion}
\end{equation}
where we defined the integer $k'$ as $k' = N\omega_a \Delta t/2 \pi = T_\mathrm{int}\omega_a /2 \pi$ and used the trigonometric expansion of the cosine of two angles. Rearranging the RHS of the equation above,
\begin{equation}
    \begin{split}
      \sum_{m,m'=0}^{N-1} (c_{m,k'}c_{m',k'}+s_{m,k'}s_{m',k'})c_{m,k}c_{m',k} & = \sum_{m'=0}^{N-1}c_{m',k'}c_{m',k} \sum_{m=0}^{N-1}c_{m,k'}c_{m,k} \\ & \quad +  \sum_{m'=0}^{N-1}s_{m',k'}c_{m',k} \sum_{m=0}^{N-1}s_{m,k'}c_{m,k} \\
      &= \left(\frac{N}{2}\right)^2\delta_{k',k} \,,
    \end{split}
    \label{eq:app:TrigIndentity}
\end{equation}
where the second equality follows from the orthogonality relations of trigonometric functions. Using this result and expressing the sum in terms of frequency indices, Eq.~\eqref{eq:app:RDMRDM} can be simplified as
\begin{equation}
\begin{split}
     \left\langle R^{(i,j)}_{k} R^{(i',j')}_{k}\right\rangle &= \frac{(\Delta t)^2}{T_\mathrm{int}} \frac{\Delta z^{(i,j)}}{L} \frac{\Delta z^{(i',j')}}{L} \\
     &\quad \times\sum_{k'} F_\mathrm{DM}(v_{k'}) A_{k'}^2 \left(\frac{N}{2}\right)^{2} \delta_{k',k} \, , \\
     &= \frac{\pi}{2} \frac{\Delta z^{(i,j)}}{L} \frac{\Delta z^{(i',j')}}{L} \frac{F_\mathrm{DM}(v_{k})}{\Delta \omega} A_{k}^2 \, ,
\end{split}
\label{eq:app:RDMRDMFirstStep}
\end{equation}
where in going to the second line we used the property of the Kronecker delta and the definition of the experimental angular frequency resolution $\Delta \omega = 2 \pi/T_\mathrm{int}$. Finally, writing the integral $F_\mathrm{DM}(v_k)$ in terms of frequencies, we find
\begin{equation}
\begin{split}
    \left\langle R^{(i,j)}_{k} R^{(i',j')}_{k}\right\rangle  &= \frac{\pi}{2} \frac{\Delta z^{(i,j)}}{L} \frac{\Delta z^{(i',j')}}{L} A_{k}^2 \\
    &\quad  \times \frac{1}{\Delta \omega} \int_{\omega_k - \Delta \omega/2}^{\omega_k + \Delta \omega/2} d\omega \, \frac{f_\mathrm{DM}(v_\omega)}{m_\phi \, v_\omega} \, ,
    \label{eq:app:PSDULDMFiniteResolution}
    \end{split}
\end{equation}
where we defined $v_\omega = \sqrt{2\omega/m_\phi-2}$. 
For completeness, we define the PSD of the ULDM signal common to all gradiometer pairs in terms of all relevant experimental and phenomenological parameters as
\begin{equation}
\begin{split}
    \left \langle S_{\mathrm{DM},k} \right \rangle & = 64\pi \, \frac{\overline{\Delta \omega_A}^2}{\omega_k^2} \sin^2\left[\frac{\omega_k n L}{2}\right] \\
    & \quad \times \sin^2 \left[\frac{\omega_k(T-(n-1)L)}{2} \right]  \sin^2\left[\frac{\omega_k T}{2}\right]  \, \\
    & \quad \times \frac{1}{\Delta \omega} \int_{\omega_k - \Delta \omega/2}^{\omega_k + \Delta \omega/2} d\omega \, \frac{f_\mathrm{DM}(v_\omega)}{m_\phi \, v_\omega}\,,
\end{split}
\label{eq:app:PSDULDMContinuum}
\end{equation}
where the amplitude of the ULDM-induced oscillation in the transition frequency $\overline{\Delta \omega_A}$ is defined in Eq.~\eqref{eq: delta omega_A}. 
As a result of these arguments, it follows that Eq.~\eqref{eq:app:RDMRDMFirstStep} can be expressed as
\begin{equation}
    \left\langle R^{(i,j)}_{k} R^{(i',j')}_{k}\right\rangle = \frac{1}{2} \frac{\Delta z^{(i,j)}}{L} \frac{\Delta z^{(i',j')}}{L} \left \langle S_{\mathrm{DM}, k} \right \rangle \, .
\end{equation}

Having determined the form of the covariance matrix entry involving only the real part of the data's DFT, we proceed with computing the terms involving only the imaginary components using a simple trick. As pointed out in Ref.~\cite{Foster:2020fln}, the result for the terms involving only the imaginary components is equivalent to that involving only the real part under the redefinition $c_{m, k} \rightarrow s_{m, k}$ and  $c_{m', k}\rightarrow s_{m', k} $. Since Eq.~\eqref{eq:app:TrigIndentity} holds under this redefinition, we also find that
\begin{equation}
    \left\langle I^{(i,j)}_{k} I^{(i',j')}_{k}\right\rangle \equiv \frac{1}{2} \frac{\Delta z^{(i,j)}}{L} \frac{\Delta z^{(i',j')}}{L} \left \langle S_{\mathrm{DM}, k} \right \rangle \, .
\end{equation}

Similarly, we can determine the terms in the covariace matrix involving the product of the imaginary and real components of the data's DFT by using the mapping $c_{m, k} \rightarrow c_{m, k}$ and  $c_{m', k}\rightarrow s_{m', k} $. Differently from the other products, we now find a sum over a product of two cosines and one sine. Since the sine function has odd parity and the cosine is of even parity, the sum will vanish for all values of $k$:
\begin{equation}
    \left\langle R^{(i,j)}_{k} I^{(i',j')}_{k}\right\rangle \equiv 0 \, .
\end{equation}

In summary, we find that the entries in the ULDM signal covariance matrix take the form
\begin{align}
    \left\langle R^{(i,j)}_{k} R^{(i',j')}_{k}\right\rangle &= \frac{1}{2} \frac{\Delta z^{(i,j)}}{L} \frac{\Delta z^{(i',j')}}{L} \left \langle S_{\mathrm{DM}, k} \right \rangle \, , \\
    \left\langle I^{(i,j)}_{k} I^{(i',j')}_{k}\right\rangle &= \frac{1}{2} \frac{\Delta z^{(i,j)}}{L} \frac{\Delta z^{(i',j')}}{L} \left \langle S_{\mathrm{DM}, k} \right \rangle \, , \\
    \left\langle R^{(i,j)}_{k} I^{(i',j')}_{k}\right\rangle &= 0\,.
\end{align}

\subsubsection{Behaviour under long and short integration times} \label{app:ULDMCovarianceScaling}

In light of Eq.~\eqref{eq:app:PSDULDMFiniteResolution}, the non-vanishing entries of the ULDM signal covariance matrix scale differently with the integration time depending on the frequency resolution attained by the experimentalist. In the regime where the integration time $T_\mathrm{int}$ is much larger than the coherence time of the signal $\tau_c$, the experimentalist would be able to resolve the full frequency width of the ULDM signal. If we take the limit $T_\mathrm{int} \rightarrow \infty$, then the width of the frequency bins tends to zero, i.e.\ $\Delta \omega \rightarrow 0$, such that 
\begin{equation}
    \lim_{\Delta \omega \rightarrow 0} \frac{1}{\Delta \omega} \int_{\omega_k - \Delta \omega/2}^{\omega_k + \Delta \omega/2} d\omega \, \frac{f_\mathrm{DM}(v_\omega)}{m_\phi \, v_\omega} = \frac{f_\mathrm{DM}(v_{\omega_k})}{m_\phi \, v_{\omega_k}} \, ,
\end{equation}
where $\omega_k = 2\pi k/T_\mathrm{int}$. Hence, in agreement with Ref.~\cite{Foster:2017hbq, Badurina:2021lwr}, the amplitude of the PSD would be independent of the integration time. 

In the regime where the integration time $T_\mathrm{int}$ is shorter than the coherence time of the signal $\tau_c$, the experimentalist would not be able to resolve the full frequency spread of the ULDM signal. In the regime where $T_\mathrm{int} \ll \tau_c$, this further implies that the ULDM signal would be contained in one frequency bin. In this case, Eq.~\eqref{eq:app:PSDULDMFiniteResolution} would be evaluated over the entire dark matter speed distribution. This implies that the integral in Eq.~\eqref{eq:app:PSDULDMContinuum} would be equal to one. Thus, recalling that $\Delta \omega = 2\pi/T_\mathrm{int}$ and that $\left \langle S_{\mathrm{DM}, k} \right \rangle \propto 1/\Delta \omega$, we find that the expected ULDM signal PSD would scale linearly with the integration time, which agrees with Refs.~\cite{Foster:2017hbq, Badurina:2021lwr}.

\subsection{Rayleigh wave covariance matrix}

The GGN covariance matrix can be calculated by recalling the statistical properties of Rayleigh waves. As for the ULDM signal, we shall first focus on the covariance between the real parts of the data's DFT, 
\begin{equation}
\begin{split}
    \left\langle R^{(i,j)}_{k} R^{(i',j')}_{k}\right\rangle &= \frac{(\Delta t)^2}{T_\mathrm{int}} \sum_{m,m'=0}^{N-1}\left \langle \Phi^{(i,j)}_{\mathrm{GGN},m}\Phi^{(i',j')}_{\mathrm{GGN},m'}\right \rangle c_{m, k}\, c_{m', k} \\
    &= \frac{(\Delta t)^2}{T_\mathrm{int}} \sum_{m,m'=0}^{N-1} \sum_{a} \frac{\langle \xi_a^2 \rangle}{2} F_{\mathrm{GGN},a}^{(i,j)}F_{\mathrm{GGN},a}^{(i',j')} \\ & \qquad \qquad \qquad \qquad  \cos (\omega_a \Delta t(m-m'))   \, c_{m, k}\, c_{m', k}\, ,
\end{split}
\label{eq:app:RGGNRGGN1}
\end{equation}
where the second equality follows from the covariance of the GGN gradiometer phase, shown in Eq.~\eqref{eq:app:GGNCovariance}. We remind the reader that we use the convention  
\begin{gather}
        F_{\mathrm{GGN},a}^{(i,j)} = F_{\mathrm{GGN},a}^{(i)}-F_{\mathrm{GGN},a}^{(j)} \, , \\ F_{\mathrm{GGN},a}^{(i)} = \widetilde{A}_a \exp \left (-q\frac{\omega_a z_i}{c_H} \right) + \widetilde{B}_a \exp \left ( -\frac{\omega_a z_i}{c_H} \right) \, ,
\end{gather}
which was presented in the main body of the paper. Since the statistics of the ULDM signal are equivalent to those of the GGN gradiometer phase, we follow the calculations in Appendix~\ref{app:CovarianceMatrices} to determine a simplified analytical form of Eq.~\eqref{eq:app:RGGNRGGN1}. Specifically, following the steps in Eqs.~\eqref{eq:app:CosineExpansion}-\eqref{eq:app:TrigIndentity}, we may express the RHS of Eq.~\eqref{eq:app:RGGNRGGN1} as
\begin{equation}
\begin{aligned}
    \left\langle R^{(i,j)}_{k} R^{(i',j')}_{k}\right\rangle &=\frac{\pi}{2} \frac{1}{\Delta \omega} \frac{\left \langle \xi_k^2 \right \rangle}{2} F_{\mathrm{GGN},k}^{(i,j)}F_{\mathrm{GGN},k}^{(i',j')}\,.
\end{aligned}
\end{equation}
To make contact with the power spectral density of vertical displacements $S_{\xi}(\omega)$ (see Fig.~\ref{fig: PSD vertical displacement}), we relate the effective mean-squared vertical displacement at angular frequency $\omega_k$ to $\langle S_{\xi}(\omega)\rangle$ via 
\begin{equation}
\begin{aligned}
    \left \langle \xi_k^2 \right \rangle &= \int_{\omega_k - \Delta \omega/2}^{\omega_k + \Delta \omega/2} d\omega \, \left \langle S_\xi(\omega) \right \rangle \, . 
\end{aligned}
\end{equation}
In light of the fact that the spectrum of vertical displacements varies significantly over frequency scales greater than $\sim 10^{-2}$~Hz and that $\Delta \omega = 2\pi/T_\mathrm{int} \approx \SI{6e-6}{Hz}$ in the experiments considered here, we may treat $S_{\xi}(\omega)$ as being constant within the integral, such that  
\begin{equation}
\begin{aligned}
    \left\langle R^{(i,j)}_{k} R^{(i',j')}_{k}\right\rangle &=\frac{\pi}{4} \left \langle S_\xi(\omega_k) \right \rangle F_{\mathrm{GGN},k}^{(i,j)}F_{\mathrm{GGN},k}^{(i',j')}\,.
\end{aligned}
\end{equation}

Following the same mapping arguments presented for the entries of the ULDM signal covariance matrix, the entries in the GGN covariance matrix take the form
\begin{align}
    \left\langle R^{(i,j)}_{k} R^{(i',j')}_{k}\right\rangle &= \frac{\pi}{4} \left \langle S_\xi(\omega_k) \right \rangle F_{\mathrm{GGN},k}^{(i,j)}F_{\mathrm{GGN},k}^{(i',j')}\, , \\
    \left\langle I^{(i,j)}_{k} I^{(i',j')}_{k}\right\rangle &=\frac{\pi}{4} \left \langle S_\xi(\omega_k) \right \rangle F_{\mathrm{GGN},k}^{(i,j)}F_{\mathrm{GGN},k}^{(i',j')}\,  , \\
    \left\langle R^{(i,j)}_{k} I^{(i',j')}_{k}\right\rangle &= 0 \, .
\end{align}

\subsection{Atom shot noise covariance matrix}

The atom shot noise covariance matrix can be calculated by recalling the statistical properties of white noise. Using the statistical properties of atom shot noise as presented in Eq.~\eqref{eq: atom shot noise covariance}, we find that the covariance between the real parts of the data's DFT takes the form
\begin{equation}
    \begin{split}
    \left\langle R^{(i,j)}_{k} R^{(i',j')}_{k}\right\rangle &= \frac{(\Delta t)^2}{T_\mathrm{int}} \sum_{m,m'=0}^{N-1}\left \langle \Phi^{(i,j)}_{\mathrm{ASN},m}\Phi^{(i',j')}_{\mathrm{ASN},m'}\right \rangle c_{m, k} c_{m', k} \\
    &= \frac{(\Delta t)^2}{T_\mathrm{int}} \sum_{m,m'=0}^{N-1} \left ( \delta^{ii'} +\delta^{jj'}-\delta^{ij'}-\delta^{ji'} \right )\delta_{m m'} \sigma_{\mathrm{Atom}}^2 c_{m, k} c_{m', k} \\
    &= \frac{(\Delta t)^2}{T_\mathrm{int}} \left ( \delta^{ii'} +\delta^{jj'}-\delta^{ij'}-\delta^{ji'} \right ) \sigma_{\mathrm{Atom}}^2 \sum_{m=0}^{N-1}  c_{m, k}^2 \,,
    \end{split}
\end{equation}
where in going to the last line we performed the sum over discrete frequencies and used the property of the Kronecker delta. Recalling the orthogonality relation for cosines, we find 
\begin{equation}  
     \left\langle R^{(i,j)}_{k} R^{(i',j')}_{k}\right\rangle= \frac{\Delta t}{2} \sigma_{\mathrm{Atom}}^2 \left ( \delta^{ii'} +\delta^{jj'}-\delta^{ij'}-\delta^{ji'} \right)\,.
    \label{eq:app:RAtomRAtom}
\end{equation}
By defining the PSD of atom shot noise in a single interferometer as $\langle S_{\mathrm{ASN},k} \rangle = \Delta t \sigma_\mathrm{Atom}^2$, we can re-express Eq.~\eqref{eq:app:RAtomRAtom} as
\begin{equation}  
     \left\langle R^{(i,j)}_{k} R^{(i',j')}_{k}\right\rangle= \frac{1}{2} \langle S_{\mathrm{ASN},k} \rangle \left ( \delta^{ii'} +\delta^{jj'}-\delta^{ij'}-\delta^{ji'} \right)\,.
    \label{eq:app:RAtomRAtom2}
\end{equation}
Following the same mapping arguments that we presented for the ULDM signal covariance matrix, we find that the entries in the atom shot noise covariance matrix take the form
\begin{align}
    \left\langle R^{(i,j)}_{k} R^{(i',j')}_{k}\right\rangle &= \frac{1}{2}\langle S_{\mathrm{ASN},k} \rangle \left ( \delta^{ii'} +\delta^{jj'}-\delta^{ij'}-\delta^{ji'} \right ) \, , \\
    \left\langle I^{(i,j)}_{k} I^{(i',j')}_{k}\right\rangle &= \frac{1}{2} \langle S_{\mathrm{ASN},k} \rangle \left ( \delta^{ii'} +\delta^{jj'}-\delta^{ij'}-\delta^{ji'} \right ) \, , \\
    \left\langle R^{(i,j)}_{k} I^{(i',j')}_{k}\right\rangle &= 0 \, .
\end{align}

\section{Asimov test statistics \label{app:AsimovTestStatistics}}

In section~\ref{sec:StatsFormalism} we defined the basic frequentist tool in the Asimov approach as
\begin{equation}
    \widetilde{\Theta} (\boldsymbol{\theta}_{\mathrm{sig}}) = \langle \Theta ( \boldsymbol{\theta}_{\mathrm{sig}}) \rangle \, ,
\end{equation}
where $ \Theta ( \boldsymbol{\theta}_{\mathrm{sig}})$ is defined in Eq.~\eqref{eq: log profile likelihood} and the average is taken over data realizations. Recalling that we have assumed a covariance matrix $\boldsymbol{\mathrm{\Sigma}}_{k}$ that can be split into signal $\boldsymbol{\mathrm{S}}_{k}$ and noise $\boldsymbol{\mathrm{B}}_{k}$ contributions, the Asimov test statistic can be expressed as 
\begin{equation}
    \left \langle \Theta\left(\boldsymbol{\theta}_{\mathrm{sig}}\right) \right \rangle = \left \langle \sum_{k=1}^{N-1}\left (\mathbf{d}_k^T \left [\mathbf{B}^{-1}_k-\mathbf{\Sigma}^{-1}_k \right]\mathbf{d}_k-\mathrm{ln}\left [\frac{|\boldsymbol{\Sigma}_k|}{|\mathbf{B}_k|}\right] \right) \right \rangle \,,
    \label{eq: TS covariances}
\end{equation}
where $\mathbf{d}_k$ is the multi-dimensional data vector. Because the entries of each covariance matrix correspond to expectation values, the average is taken only over the first term in Eq.~\eqref{eq: TS covariances}. Since the data has zero mean, we can define $\mathbf{\Sigma}^t_k \equiv \left \langle \mathbf{d}_k \mathbf{d}_k^T \right \rangle$ as the true covariance matrix, such that
\begin{equation}
    \left \langle \mathbf{d}_k^T \left [\mathbf{B}_k^{-1}-\mathbf{\Sigma}_k^{-1}\right]\mathbf{d}_k \right \rangle = \mathrm{Tr} \left ( \mathbf{\Sigma}^t_k \left [\mathbf{B}_k^{-1}-\mathbf{\Sigma}_k^{-1}\right]  \right) \, .
\end{equation}
Using this result and recalling the trace-determinant identity for symmetric matrices, Eq.~\eqref{eq: TS covariances} takes the form
\begin{align}
      \widetilde{\Theta}\left(\boldsymbol{\theta}_{\mathrm{sig}}\right) &= \sum_{k=1}^{N-1} \mathrm{Tr}\left ( \mathbf{\Sigma}^t_k \left[\mathbf{B}^{-1}_k-\mathbf{\Sigma}^{-1}_k \right]+\mathrm{ln}\left [\mathbf{B}_k\mathbf{\Sigma}_k^{-1} \right]   \right)\,. 
      \label{eq: TS final}
\end{align}
Given a fixed background, the true covariance matrix can be defined as $\mathbf{\Sigma}^t_k = \mathbf{S}^t_k + \mathbf{B}_k$, where $\mathbf{S}^t_k$ is the true signal model. In the limit that the signal is weaker than the background, i.e.\ $\mathbf{\Sigma}_k^{-1} \approx \mathbf{B}_k^{-1}-\mathbf{B}_k^{-1}\mathbf{S}_k\mathbf{B}_k^{-1}$, we can expand the natural logarithm of the matrix product close to the identity matrix such that
\begin{align}
      \widetilde{\Theta}\left(\boldsymbol{\theta}_{\mathrm{sig}}\right) &\approx \sum_{k=1}^{N-1} \mathrm{Tr}\left [ \left ( \mathbf{S}^t_k-\frac{1}{2}\mathbf{S}_k\right ) \mathbf{B}^{-1}_k \mathbf{S}_k \mathbf{B}^{-1}_k \right] \, .
      \label{eq: TS approx}
\end{align}

For setting upper limits on the ULDM-SM couplings~$d_\phi$, we use the test statistic $q\left(m_\phi, d_\phi\right)$. The equivalent test statistic on the Asimov data set can then be defined as 
\begin{equation}
    \widetilde{q}\left(m_{\phi}, d_\phi\right)= \begin{cases} \widetilde{\Theta}\left(\{m_{\phi}, d_\phi\right\})-\widetilde{\Theta}\left(\{m_{\phi}, \hat{d_\phi}\}\right) & d_\phi \geq \hat{d_\phi} \\ 0 & d_\phi<\hat{d_\phi}\end{cases}\;,
\end{equation}
where $\hat{d_\phi}$ is the value of $d_\phi$ that maximises $\widetilde{\Theta}\left(m_\phi, d_\phi\right)$ at fixed $m_\phi$. In this case, the appropriate Asimov dataset contains only the background, such that $\hat{d_\phi} = 0$, which implies that the true covariance matrix can be written as $\mathbf{\Sigma}^t_k = \mathbf{B}_k$ - i.e.\ $\mathbf{S}^t_k=\mathbf{0}$. Therefore, in the weak signal limit, we may express the test statistic for setting upper limits as follows:
\begin{equation}
    \widetilde{q}\left(m_{\phi}, d_\phi\right)= \begin{cases} -\frac{1}{2}\sum_{k=1}^{N-1} \mathrm{Tr}\left [\mathbf{S}_k \mathbf{B}^{-1}_k \mathbf{S}_k \mathbf{B}^{-1}_k \right]  & d_\phi \geq 0 \\ 0 & d_\phi<0\end{cases}\;.
    \label{eq:app:TSUpperLimits}
\end{equation}

\subsection{Behaviour under long and short integration times } \label{app:TestStatisticScaling}

In practice, the sum in Eq.~\eqref{eq: TS approx} is only to be performed around a small angular frequency window $m_\phi \lesssim \omega \lesssim m_\phi(1+v_\mathrm{esc}^2/2)$, where $v_\mathrm{esc}$ is the dark matter escape velocity in the Milky Way. In the regime where $T_\mathrm{int}$ is the longest timescale, i.e.\ $T_\mathrm{int} \gg \tau_c$, the sum over the relevant frequency bins can be approximated to an integral over frequency indices, such that, for $d_\phi\geq 0$, Eq.~\eqref{eq:app:TSUpperLimits} is approximately
\begin{equation}
\begin{aligned}
    \widetilde{q}\left(m_{\phi}, d_\phi\right) & \approx -\frac{1}{2} \int dk \, \mathrm{Tr}\left [\mathbf{S}_k \mathbf{B}^{-1}_k \mathbf{S}_k \mathbf{B}^{-1}_k \right] \\ 
    & \sim \int dk \left \langle R_{\mathrm{DM},k}^2 \right \rangle^2/ \left \langle R_{\mathrm{Noise},k}^2 \right  \rangle^2 \, .
\end{aligned}
\end{equation}
In addition, since the normalised frequency band over which the sum is performed is $\sim 10^{-6}$ and GGN is not expected to vary appreciably over this normalised frequency range, we treat all backgrounds as being constant. Hence, recalling the relation $2\pi dk/ T_\mathrm{int}= m_\phi v dv$, we find 
\begin{equation}
\begin{aligned}
    \widetilde{q}\left(m_{\phi}, d_\phi\right) & \sim \frac{1}{2\pi}\frac{m_\phi T_\mathrm{int}}{\langle R^2_\mathrm{Noise}\rangle^2}\int v\, dv \, \left \langle R_{\mathrm{DM},v}^2 \right \rangle^2 \\
    & \propto \frac{T_\mathrm{int}}{\langle R^2_\mathrm{Noise}\rangle^2}\int \frac{dv}{m_\phi v} \, f^2_\mathrm{DM}(v) \, ,
\end{aligned}
\end{equation}
where in going to the second line we used $ \langle R_{\mathrm{DM},v}^2 \rangle \sim f_\mathrm{DM}(v)/m_\phi v$, which follows from the discussion in Appendix~\ref{app:ULDMCovarianceScaling}. Because $f_\mathrm{DM}$ is a Gaussian distribution and since $v_\mathrm{obs} \sim v_0$, the integral in the last line of this equation is approximately equal to $1/m_\phi v_0^2$. Recalling the definition of the coherence time $\tau_c = 2\pi/(m_\phi v_0^2)$, we find
\begin{equation}
\widetilde{q}\left(m_{\phi}, d_\phi\right) \propto T_\mathrm{int}\tau_c \, .
\end{equation}
Hence, as derived via Bartlett's method in Ref.~\cite{Badurina:2021lwr}, the sensitivity to ULDM would scale like $(T_\mathrm{int} \tau_c)^{-1/4}$.

In the regime where $T_\mathrm{int} \ll \tau_c$, the experiment is unable to resolve the DM speed distribution. In this case, the signal is confined to a single frequency bin, such that, for $d_\phi\geq 0$, Eq.~\eqref{eq:app:TSUpperLimits} is approximately
\begin{equation}
\begin{aligned}
    \widetilde{q}\left(m_{\phi}, d_\phi\right) & \approx -\frac{1}{2}  \mathrm{Tr}\left [\mathbf{S}_k \mathbf{B}^{-1}_k \mathbf{S}_k \mathbf{B}^{-1}_k \right] \\ 
    & \sim \left \langle R_{\mathrm{DM},k}^2 \right \rangle^2/ \left \langle R_{\mathrm{Noise},k}^2 \right \rangle^2 \, .
\end{aligned}
\end{equation}
Following the analysis presented in Appendix~\ref{app:ULDMCovarianceScaling}, we make use of the fact that $ \langle R_{\mathrm{DM},k}^2 \rangle \propto 1/\Delta \omega \propto T_\mathrm{int}$ to determine that the test statistic for setting upper limits scales as $T_\mathrm{int}^2$. Hence, in the regime where $T_\mathrm{int} \ll \tau_c$ the sensitivity to ULDM would scale like $T_\mathrm{int} ^{-1/2}$, in agreement with Ref.~\cite{Badurina:2021lwr}. 

\subsection{Example power spectral densities}

\begin{figure*}[t!]
    \centering
    \includegraphics[width=.95\textwidth]{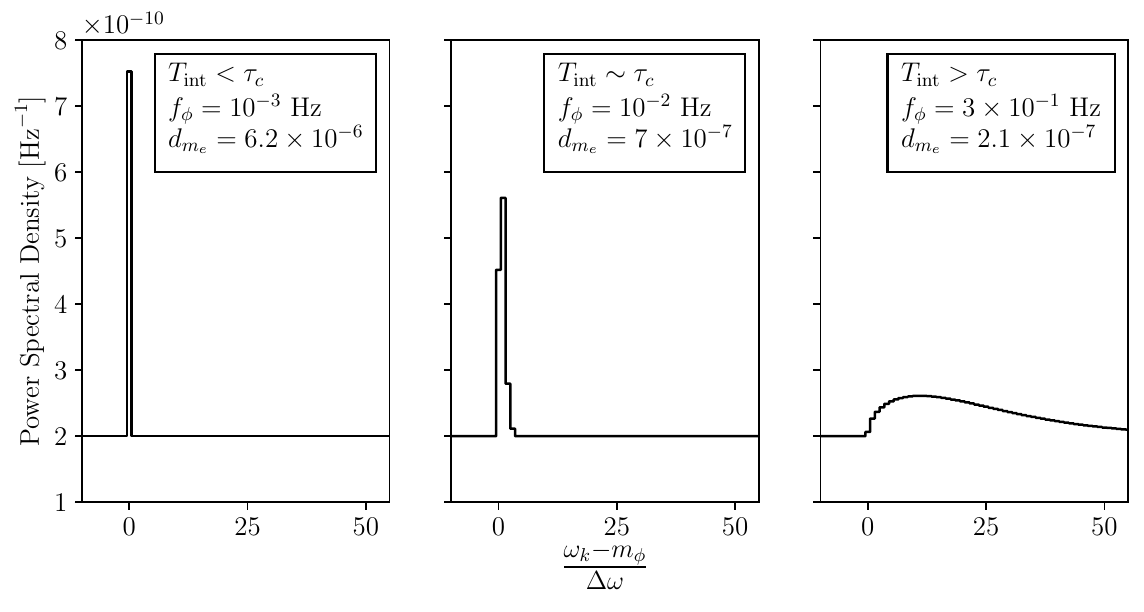}
    \caption{Example power spectral densities in a gradiometer experiment employing the `advanced' design, assuming atom shot noise only, and taking couplings and DM masses compatible with the 95\%~CL shown in the right panel of Fig.~~\ref{fig:SensitivityAION1km_N2}. In the left panel, we choose $f_\phi = 10^{-3}$~Hz and $d_{m_e} = 6.2\times 10^{-6}$ ($T_\mathrm{int} < \tau_c$); in the central panel, we choose $f_\phi = 10^{-2}$~Hz and $d_{m_e} = 7\times 10^{-7}$ ($T_\mathrm{int} \sim \tau_c$); in the right panel, we set $f_\phi = 3\times 10^{-1}$~Hz and $d_{m_e} = 2.1\times 10^{-7}$ ($T_\mathrm{int} > \tau_c$).
    } 
    \label{fig:example PSD}
\end{figure*}

In Fig.~\ref{fig:example PSD} we display the expected power spectral density in a gradiometer experiment employing the `advanced' design ($T_\mathrm{int} = 10^8$~s) and assuming atom shot noise only in three different regimes: $T_\mathrm{int} < \tau_c$ (left panel), $T_\mathrm{int} \sim \tau_c$ (central panel) and $T_\mathrm{int} > \tau_c$ (right panel). In particular, we show the expected signal assuming couplings to electron mass ($d_{m_e}$) and ULDM frequencies ($f_\phi$) that lie on the grey dashed line in the right panel of Fig.~\ref{fig:SensitivityAION1km_N2}, and which therefore are compatible with the 95\% exclusion limits: $f_\phi = 10^{-3}$~Hz and $d_{m_e} = 6.2\times 10^{-6}$ (left panel), $f_\phi = 10^{-2}$~Hz and $d_{m_e} = 7\times 10^{-7}$ (central panel), and $f_\phi = 3\times 10^{-1}$~Hz and $d_{m_e} = 2.1\times 10^{-7}$ (right panel). 

As discussed in Appendix~\ref{app:ULDMcovariancematrix}, when $T_\mathrm{int} > \tau_c$, the signal is spread over many frequency bins; in particular, for $T_\mathrm{int} \gg \tau_c$, the signal clearly exhibits the qualitative features of the DM speed distribution. In the regime $T_\mathrm{int} \sim \tau_c$, the signal is spread over few bins. For $T_\mathrm{int} \ll \tau_c$, the signal is contained within a single frequency bin centred at about $m_\phi$, which we display as the zeroth frequency bin in Fig.~\ref{fig:example PSD}. 

Additionally, as shown in Fig.~\ref{fig:example PSD}, for a fixed value of $q$ the maximum amplitude of the signal's PSD decreases as $m_\phi$ increases. This is attributable to the properties of the test statistic $q$. In setting upper limits, the test statistic $q$ receives non-vanishing contributions when the PSD of the ULDM signal is non-zero. Hence, the larger the signal's frequency spread (equivalently the larger $m_\phi$), the larger the number of bins contributing to $q$. For a fixed threshold value of $q$, an increase in the number of frequency bins, hence, implies a smaller signal amplitude.  

In the extreme case when the ULDM signal is contained in a single frequency bin ($T_\mathrm{int} \ll \tau_c$), the $q$ test statistic is a measure of the  SNR as defined in Refs.~\cite{ Badurina:2021rgt, Arvanitaki:2016fyj}, namely the ratio of the signal's PSD to the background's PSD. In this regime, the equality $\mathrm{SNR} = \sqrt{q}$ holds exactly. As can be seen from the left panel in Fig.~\ref{fig:example PSD}, for $q \approx 7.55 $, the signal PSD at the zeroth frequency bin ($\approx 5.5 \times 10^{-10}$~Hz\textsuperscript{-1}) is about $\sqrt{7.55}$ times larger than the noise PSD ($\approx 2 \times 10^{-10}$~Hz\textsuperscript{-1}). 

\bibliography{ref}
\bibliographystyle{JHEP}
\end{document}